\documentclass[12pt]{JHEP3}

\usepackage{amsmath}
\usepackage{epsfig}
\usepackage{epstopdf}
\usepackage{graphicx}

\numberwithin{equation}{section} \numberwithin{table}{section}
\numberwithin{figure}{section}

\title{Lovelock-Lifshitz Black Holes}
\author{M. H. Dehghani\\
1. Department of Physics, University of Waterloo, 200 University
Avenue West,\\ \ \ \ Waterloo, Ontario, Canada, N2L 3G1\\
2. Physics Department and Biruni Observatory,
College of Sciences,\\ \ \ \ Shiraz University, Shiraz 71454, Iran\\
E-mail: \email{dehghani@uwaterloo.ca, mhd@shirazu.ac.ir}}
\author{R. B. Mann\\ Department of Physics, University of Waterloo, 200 University
Avenue West, Waterloo, Ontario, Canada, N2L 3G1\\
E-Mail: \email{rbmann@sciborg.uwaterloo.ca}
}
\abstract{
In this paper, we investigate the existence
of Lifshitz solutions in Lovelock gravity, both in vacuum and in the presence of
a massive vector field.  We show that the Lovelock terms
can support the Lifshitz solution provided the constants of the theory are suitably chosen.
We obtain an exact black hole solution with Lifshitz asymptotics of any scaling parameter $z$
in both Gauss-Bonnet and in pure 3rd order Lovelock gravity.
If matter is added in the form of a massive vector field,  we also show that  Lifshitz solutions in Lovelock gravity exist; these can be regarded as corrections to Einstein gravity coupled to this
form of matter.  For this form of matter we numerically obtain a broad range of charged black hole solutions with Lifshitz asymptotics, for either sign of the cosmological constant.   We find that these
asymptotic Lifshitz solutions are more
sensitive to  corrections induced by Lovelock gravity than are their asymptotic
AdS counterparts. We also consider the thermodynamics of the black hole solutions
and show that the temperature of large black holes with curved horizons is proportional to $r_0^z$ where $z$ is the critical exponent;  this relationship holds for black branes of any size.
As is the case for asymptotic AdS
black holes, we find that an extreme black hole exists only for the case of horizons with negative
curvature.  We also find that these Lovelock-Lifshitz black holes have no unstable phase, in contrast to the Lovelock-AdS case.   We also present a class of rotating Lovelock-Lifshitz black holes with
Ricci-flat horizons.}

\begin{document}
\tableofcontents
\section{Introduction}

Candidates for gravity duals of non-relativistic scale invariant
theories have recently attracted a great deal of attention for several reasons.
One is that some condensed matter systems realized in laboratories at their
critical points are described by non-relativistic conformal field theories.
Furthermore, the AdS/CFT correspondence \cite{Mal,Wit} describing the duality
between strongly coupled conformal field theory and gravity continues to
find more applications in other branches of physics such as QCD quark-gluon
plasmas \cite{Kov}, atomic physics, and condensed matter physics \cite
{Hart,Hart2,Hart3,Herz,Faul,Mc}. This has led to a study of gravity-gauge duality in a
much broader context than its original AdS/CFT formulation, extending to
non-relativistic and Lifshitz field theories, with the aim of gaining more
knowledge of the strong coupling behavior of these other physical
 systems.
Non-relativistic conformal symmetry  contains the scaling symmetry
\begin{equation}
t\rightarrow \lambda ^{z}t,\hspace{1cm}\mathbf{{x}\rightarrow \lambda {x}}
\label{Ssym}
\end{equation}
where z is the dynamical exponent. This transformation exhibits anisotropic
scale invariant behavior. Actually, for condensed matter applications, one
typically is interested in considering anisotropy between different spatial
dimensions. For $z=1$ this scaling symmetry is the familiar conformal
symmetry. Such a non-relativistic scale invariance (\ref{Ssym}) can be
exhibited by either a Galilean-invariant theory or a Lifshitz-invariant
theory. From a holographic point of view, this suggests the following (asymptotic)
form for the
spacetime metric
\begin{equation}
ds^{2}=L^{2}\left( -r^{2z}dt^{2}+\frac{dr^{2}}{r^{2}}+r^{2}d\mathbf{x}%
^{2}\right)  \label{Lifmet}
\end{equation}
that obeys the scale invariance
\begin{equation}
t\rightarrow \lambda ^{z}t,\hspace{0.5cm}r\rightarrow \lambda ^{-1}r,\hspace{%
0.5cm}\mathbf{{x}\rightarrow \lambda {x}}  \label{Ssym2}
\end{equation}
noted previously in a braneworld context  \cite{Koroteev}.

A four-dimensional anisotropic scale invariant background using an action
involving a two form and a three form field with a Chern-Simons coupling
\begin{equation}\label{actHB}
I=\int d^{4}x\sqrt{-g}\left( R-2\Lambda -\frac{1}{4}F_{\alpha \beta}F^{\mu
\nu } -\frac{1}{12}H_{\mu \nu \rho}H^{\mu \nu \rho} -\frac{C}{\sqrt{-g}}
F\wedge B\right),
\end{equation}
can be engineered to yield solutions with this asymptotic behavior \cite
{Kach}, where $H=dB$ and $F=dA$. Such an action was argued to be rather
generic in string theory, although no explicit brane realization or
embedding into ten-dimensions was given. For these matter fields, lots of
effort has been expended in extending this solution to the case of
asymptotic Lifshitz solutions. One of the first analytic examples was
reported in Ref. \cite{Tay} for a sort of higher-dimensional dilaton gravity
without restricting the value of the dynamical exponent $z$. An exact
topological black hole solution with hyperbolic horizon which happens to be
asymptotically Lifshitz with $z = 2$ was found in \cite{Mann}; further
solutions with $z = 4$ and with spherical topology were subsequently
obtained \cite{Peet}. However in general such asymptotic Lifshitz black
holes must be investigated numerically \cite{Mann,Peet,Bal,Dan}. Other
possibilities for the matter needed to support such a background have been
investigated \cite{Tay}. Asymptotic Lifshitz solutions in the vacuum of
higher-derivative gravity theories (with curvature-squared terms in the
action) have been investigated \cite{Ay1,Ay2,Cai,Pang}; the higher curvature terms with
suitable coupling constant play the role of the desired matter.

Recently gravity theories including higher powers of the curvature, particularly
curvature-cubed interactions, have attracted increased attention \cite{DP}.
This is because, in the context of AdS/CFT correspondence, corrections from higher powers of the
curvature must be considered on the gravity side of the correspondence in order to scrutinize CFTs with different values for their central charges.  Amongst the cornucopia of higher-curvature gravity theories, Lovelock gravity theories play a special role in that the number of metric
derivatives in any field equation  is never larger than 2.  Third-order Lovelock gravity is supersymmetric, and therefore one can define superconformal field theories via the AdS/CFT
correspondence \cite{Boer,Cam}.  Furthermore quasi-topological gravity  including
curvature-cubed interactions, while not supersymmetric,  can
be considered to be dual to some non-supersymmetric but
conformal gauge theory in the limit of a large number of colours  \cite{Myers}.

In this paper, we
consider the existence of Lifshitz solutions in third order Lovelock gravity both in
vacuum and in the presence of a massive vector field. Since the higher
curvature terms appear to play the role of some kind of matter field, it is
natural to ask whether they can support a Lifshitz solution in vacuum or
not. We find that the answer is yes, albeit under restricted circumstances.
We also search for asymptotic Lifshitz black holes in the presence of a
massive vector field, whose action is given via a dualization of the action
(\ref{actHB}). The solutions we find -- both analytically and
numerically -- can be regarded as higher-curvature modifications to those
obtained from Einsteinian gravity coupled to matter \cite{Mann,Peet,Dan}.

The outline of our paper is as follows. We give a brief review of the field
equations of third order Lovelock gravity in the presence of massive vector
field in Sec. \ref{Fiel}. In Sec. \ref{Lif} we present the $(n+1)$%
-dimensional exact Lifshitz solutions in vacuum and in the presence of a
massive vector field. In Sec. \ref{black} we obtain the series solutions to
the field equations near the horizon, while the series solutions at large $r$
will be given the Appendix. We then obtain
numerical solutions to these equations. The thermodynamics of these
Lovelock-Lifshitz black holes will be given in Sec. \ref{Therm}. Section \ref
{rot} will be devoted to the rotating Lovelock-Lifshitz solutions. We finish
our paper with some concluding remarks.

\section{Field equations\label{Fiel}}

The fundamental assumptions in standard general relativity are the
requirements of general covariance and 2nd-order differential field
equations for the metric. Based on the same principles, the Lovelock
Lagrangian is a very general Lagrangian in classical gravity that produces
second order field equations for the metric for arbitrary powers of the curvature \cite{Lov}. The action of third order Lovelock
gravity in the presence of an Abelian massive vector field $A^{\mu }$ may be
written as
\begin{equation}
I=\int d^{n+1}x\sqrt{-g}\left( -2\Lambda +\mathcal{L}_{1}+\alpha _{2}%
\mathcal{L}_{2}+\alpha _{3}\mathcal{L}_{3}-\frac{1}{4}F_{\mu \nu }F^{\mu \nu
}-\frac{1}{2}m^{2}A_{\mu }A^{\mu }\right) ,  \label{Act1}
\end{equation}
where $F_{\mu \nu }=\partial_{[\mu}A_{\nu]}$, $\Lambda $ is the cosmological
constant, $\alpha _{2}$ and $\alpha _{3} $ are Gauss-Bonnet and third order
Lovelock coefficients, $\mathcal{L}_{1}=R$ is the Einstein-Hilbert
Lagrangian, $\mathcal{L}_{2}=R_{\mu \nu \gamma \delta }R^{\mu \nu \gamma
\delta }-4R_{\mu \nu }R^{\mu \nu }+R^{2}$ is the Gauss-Bonnet Lagrangian, and

\begin{eqnarray}
\mathcal{L}_{3} &=&R^{3}+2R^{\mu \nu \sigma \kappa }R_{\sigma \kappa \rho \tau }R_{%
\phantom{\rho \tau }{\mu \nu }}^{\rho \tau }+8R_{\phantom{\mu \nu}{\sigma
\rho}}^{\mu \nu }R_{\phantom {\sigma \kappa} {\nu \tau}}^{\sigma \kappa }R_{%
\phantom{\rho \tau}{ \mu \kappa}}^{\rho \tau }+24R^{\mu \nu \sigma \kappa
}R_{\sigma \kappa \nu \rho }R_{\phantom{\rho}{\mu}}^{\rho }  \notag \\
&&+3RR^{\mu \nu \sigma \kappa }R_{\sigma \kappa \mu \nu }+24R^{\mu \nu
\sigma \kappa }R_{\sigma \mu }R_{\kappa \nu }+16R^{\mu \nu }R_{\nu \sigma
}R_{\phantom{\sigma}{\mu}}^{\sigma }-12RR^{\mu \nu }R_{\mu \nu }
\label{L3}
\end{eqnarray}
is the third order Lovelock Lagrangian. We assume that
the Gauss-Bonnet coefficient, which has the dimension of (length)$^2$, is positive as
in the heterotic string theory \cite{Boul}.

In Lovelock gravity only terms with order less than $[(n+1)/2]$ (where $[x]$
is the integer part of $x$) contribute to the field equations, the rest
being total derivatives in the action. For 3rd-order Lovelock gravity we
therefore consider $(n+1)$-dimensional spacetimes with $n\geq 6$
(though in situations where we set $\hat{\alpha}_3=0$ our solutions will
be valid for $n\geq 4$).
Varying
the action with respect to the metric tensor $g_{\mu \nu }$ and gauge field $%
A_{\mu }$ the equations of gravitation and gauge fields are
\begin{eqnarray}
&&G_{\mu \nu }^{(1)}+\alpha _{2}G_{\mu \nu }^{(2)}+\alpha _{3}G_{\mu \nu
}^{(3)}+\Lambda g_{\mu \nu }=T_{\mu \nu },  \label{Geq} \\
&&\partial _{\mu }\left( \sqrt{-g}F^{\mu \nu }\right) =m^{2}\sqrt{-g}A^{\nu
},  \label{EMeq}
\end{eqnarray}
where
\begin{equation*}
T_{\mu \nu }=\frac{1}{2}\left( F_{\phantom{\lambda}{\mu}}^{\rho }F_{\rho \nu
}-\frac{1}{4}F_{\rho \sigma }F^{\rho \sigma }g_{\mu \nu }+m^{2}\left[ A_{\mu
}A_{\nu }-\frac{1}{2}A_{\lambda }A^{\lambda }g_{\mu \nu }\right] \right)
\end{equation*}
is the energy-momentum tensor of gauge field, $G_{\mu \nu }^{(1)}$ is just
the Einstein tensor, and $G_{\mu \nu }^{(2)}$ and $G_{\mu \nu }^{(3)}$ are
given as:
\begin{eqnarray*}
G_{\mu \nu }^{(2)} &=&2(-R_{\mu \sigma \kappa \tau }R_{\phantom{\kappa \tau
\sigma}{\nu}}^{\kappa \tau \sigma }-2R_{\mu \rho \nu \sigma }R^{\rho \sigma
}-2R_{\mu \sigma }R_{\phantom{\sigma}\nu }^{\sigma }+RR_{\mu \nu })-\frac{1}{%
2}\mathcal{L}_{2}g_{\mu \nu }, \\
G_{\mu \nu }^{(3)} &=&-3(4R^{\tau \rho \sigma \kappa }R_{\sigma \kappa
\lambda \rho }R_{\phantom{\lambda }{\nu \tau \mu}}^{\lambda }-8R_{%
\phantom{\tau \rho}{\lambda \sigma}}^{\tau \rho }R_{\phantom{\sigma
\kappa}{\tau \mu}}^{\sigma \kappa }R_{\phantom{\lambda }{\nu \rho \kappa}%
}^{\lambda }+2R_{\nu }^{\phantom{\nu}{\tau \sigma \kappa}}R_{\sigma \kappa
\lambda \rho }R_{\phantom{\lambda \rho}{\tau \mu}}^{\lambda \rho } \\
&&-R^{\tau \rho \sigma \kappa }R_{\sigma \kappa \tau \rho }R_{\nu \mu }+8R_{%
\phantom{\tau}{\nu \sigma \rho}}^{\tau }R_{\phantom{\sigma \kappa}{\tau \mu}%
}^{\sigma \kappa }R_{\phantom{\rho}\kappa }^{\rho }+8R_{\phantom
{\sigma}{\nu \tau \kappa}}^{\sigma }R_{\phantom {\tau \rho}{\sigma \mu}%
}^{\tau \rho }R_{\phantom{\kappa}{\rho}}^{\kappa } \\
&&+4R_{\nu }^{\phantom{\nu}{\tau \sigma \kappa}}R_{\sigma \kappa \mu \rho
}R_{\phantom{\rho}{\tau}}^{\rho }-4R_{\nu }^{\phantom{\nu}{\tau \sigma
\kappa }}R_{\sigma \kappa \tau \rho }R_{\phantom{\rho}{\mu}}^{\rho
}+4R^{\tau \rho \sigma \kappa }R_{\sigma \kappa \tau \mu }R_{\nu \rho
}+2RR_{\nu }^{\phantom{\nu}{\kappa \tau \rho}}R_{\tau \rho \kappa \mu } \\
&&+8R_{\phantom{\tau}{\nu \mu \rho }}^{\tau }R_{\phantom{\rho}{\sigma}%
}^{\rho }R_{\phantom{\sigma}{\tau}}^{\sigma }-8R_{\phantom{\sigma}{\nu \tau
\rho }}^{\sigma }R_{\phantom{\tau}{\sigma}}^{\tau }R_{\mu }^{\rho }-8R_{%
\phantom{\tau }{\sigma \mu}}^{\tau \rho }R_{\phantom{\sigma}{\tau }}^{\sigma
}R_{\nu \rho }-4RR_{\phantom{\tau}{\nu \mu \rho }}^{\tau }R_{\phantom{\rho}%
\tau }^{\rho } \\
&&+4R^{\tau \rho }R_{\rho \tau }R_{\nu \mu }-8R_{\phantom{\tau}{\nu}}^{\tau
}R_{\tau \rho }R_{\phantom{\rho}{\mu}}^{\rho }+4RR_{\nu \rho }R_{%
\phantom{\rho}{\mu }}^{\rho }-R^{2}R_{\nu \mu })-\frac{1}{2}\mathcal{L}%
_{3}g_{\mu \nu }.
\end{eqnarray*}

The metric of an $(n+1)$-dimensional asymptotically Lifshitz static
and radially symmetric
spacetime may be written as:
\begin{equation}  \label{met1}
ds^{2}=-\frac{r^{2z}}{l^{2z}}f(r)dt^{2}+\frac{l^2 dr^{2}}{r^{2}g(r)}%
+r^{2}d\Omega ^{2}
\end{equation}
where the functions $f(r)$ and $g(r)$ must go to $1$ as $r$ goes to
infinity. In Eq. (\ref{met1}) $d\Omega ^{2}$ is the metric of an $(n-1)$
-dimensional hypersurface with constant curvature $(n-1)(n-2)k$ and volume $V_{n-1}$. We
can write
\begin{equation}
d\Omega ^{2}=\left\{
\begin{array}{cc}
d\theta _{1}^{2}+\sum\limits_{i=2}^{n-1}\prod\limits_{j=1}^{i-1}\sin
^{2}\theta _{j}d\theta _{i}^{2} & k=1 \\
d\theta _{1}^{2}+\sinh ^{2}\theta _{1}\left(d\theta
_{2}^{2}+\sum\limits_{i=3}^{n-1}\prod\limits_{j=2}^{i-1}\sin ^{2}\theta
_{j}d\theta _{i}^{2}\right) & k=-1 \\
\sum\limits_{i=1}^{n-1}d\theta _{i}^{2} & k=0
\end{array}
\right.
\end{equation}
though it should be noted that our solutions are valid whenever $d\Omega^2$
describes any Einstein space.

Using the ansatz
\begin{equation}  \label{gauge}
A_{t}=q\frac{r^{z}}{l^{z}}h(r)
\end{equation}
for the gauge field and defining $\hat{\alpha }_{2}\equiv (n-2)(n-3)\alpha
_{2}$ and $\hat{\alpha }_{3}\equiv (n-2)...(n-5)\alpha _{3}$ for
convenience, the field equations (\ref{Geq}) and (\ref{EMeq}) reduce to the
system
\begin{eqnarray}
&& 2r^{2}h^{\prime \prime }- r\left[ (\ln f)^{\prime }-(\ln
g)^{\prime }\right]( rh^{\prime }+z) -2(n+z) rh^{\prime }+2(n-1)z=\frac{2m^{2}l^{2}}{g},  \label{E1} \\
&& r^{4}l^{4}\left\{ n(n-1)r^{2}g+(n-1)r^{3}g^{\prime }+2\Lambda
l^{2}r^{2}-(n-1)(n-2)kl^{2}\right\}  \notag \\
&&+(n-1)\hat{\alpha }_{2}l^{2}r^{2}(kl^{2}-r^{2}g)\left\{
nr^{2}g+2r^{3}g^{\prime }-(n-4)kl^{2}\right\}  \notag \\
&&+(n-1)\hat{\alpha }_{3}(kl^{2}-r^{2}g)^{2}\left\{ nr^{2}g+3r^{3}g^{\prime
}-(n-6)kl^{2}\right\}=2l^{6}r^{6}T_{t}^{t},  \label{E2} \\
&& r^{4}l^{4}\left\{ (n-1)(n-2+2z)r^{2}g+(n-1)r^{3}g(\ln f)^{\prime
}+2\Lambda l^{2}r^{2}-(n-1)(n-2)kl^{2}\right\}  \notag \\
&& +(n-1)\hat{\alpha }_{2}l^{2}r^{2}(kl^{2}-r^{2}g)\left\{
(n-4+4z)r^{2}g+2r^{3}g(\ln f)^{\prime }-(n-4)kl^{2}\right\} \notag \\
&&+(n-1)\hat{\alpha }_{3}(kl^{2}-r^{2}g)^{2}\left\{
(n-6+6z)r^{2}g+3r^{3}g(\ln f)^{\prime }-(n-6)kl^{2}\right\} \notag \\
&& \hspace{10cm} =2l^{6}r^{6}T_{r}^{r},  \label{E3}
\end{eqnarray}
where prime denotes the derivative with respect to $r$ and $T_{t}^{t}$ and $%
T_{r}^{r}$ are
\begin{eqnarray*}
T_{t}^{t} &=&-\frac{q^{2}}{4l^{2}f}\left\{ g(rh^{\prime }+zh)^{2}+m^2
l^2h^{2}\right\} , \\
T_{r}^{r} &=&-\frac{q^{2}}{4l^{2}f}\left\{ g(rh^{\prime }+zh)^{2}-m^2
l^2h^{2}\right\} .
\end{eqnarray*}

\section{Lifshitz Solutions\label{Lif}}

\subsection{Vacuum Solutions}

We first investigate the possibility of having $(n+1)$-dimensional Lifshitz
solutions
\begin{equation}
ds^{2}=-\frac{r^{2z}}{l^{2z}}dt^{2}+\frac{l^2 dr^{2}}{r^{2}}+r^{2}\sum%
\limits_{i=1}^{n-1}d\theta _{i}^{2},  \label{met2}
\end{equation}
in Lovelock gravity in the absence of matter. In order to have an
asymptotically Lifshitz solution in 3rd-order Lovelock gravity, the
following constraints on the cosmological constant and third order Lovelock
coefficient
\begin{equation}
\Lambda = -\frac{n(n-1)}{6l^{4}}\left( 2l^{2}-\hat{\alpha }_{2}\right) ,
\qquad \hat{\alpha }_{3} = -\frac{l^{2}}{3}(l^{2}-2\hat{\alpha }_{2}).
\label{Coef2}
\end{equation}
hold for an arbitrary value of $z$, as is easily obtained via
straightforward calculation.

Note that for $\alpha _{3}=0$ the constraints (\ref{Coef2}) become
\begin{equation}
\Lambda =-\frac{n(n-1)}{4l^{2}}\text{ \ \ and \ \ }\hat{\alpha }_{2}=\frac{%
l^{2}}{2}.  \label{Coef1}
\end{equation}
and so asymptotically Lifshitz solutions exist in the vacuum of Gauss-Bonnet
gravity ($\alpha _{3}=0$) as well. The cosmological constant here
is half that of an AdS spacetime. If we set $\Lambda=0$ then a Lifshitz
solution exists in third order Lovelock gravity provided
\begin{equation}
\hat{\alpha }_{2}=2l^{2},\text{ \ \ \ \ \ }\hat{\alpha }_{3}=l^{4}.
\label{Coef3}
\end{equation}
Thus, one may have Lifshitz solutions in Lovelock gravity without matter,
demonstrating that its higher curvature terms can,
for the proper choice of Lovelock coefficients, have the desired effect
that matter fields induce.

The preceding analysis was for $k=0$.  For $k=\pm 1$ we find for any value of
$z$ that
\begin{equation}
g(r)=1+\frac{kl^{2}}{r^{2}}  \label{exa}
\end{equation}
furnishes an exact Lifshitz solution to Lovelock gravity without matter, provided one
of the above conditions (\ref{Coef2}), (\ref{Coef1}) is satisfied, with the function $f(r)$
undetermined by the field equations (though boundary conditions constrain it to
 asymptote to 1).  We can choose $f(r)=g(r)$. This is a
naked singularity for $k=1$ but is an asymptotically Lifshitz black hole for
$k=-1$. The arbitrariness of $f(r)$ is due to a degeneracy of the field equations; if either of (\ref{Coef2}) or (\ref{Coef1}) hold then the field equations are each proportional to the factor $g(r)-1-kl^2/r^2$.

This degeneracy of the field equations has been noted previously in 5-dimensional Einstein-Gauss-Bonnet gravity with a cosmological constant \cite{Oliva}, where (converting notation appropriately)  condition  (\ref{Coef1})  was obtained.  Here we see that a more general degeneracy occurs in 3rd order Lovelock gravity.  We shall see that this degeneracy is lifted when matter is present.   We find that no other exact solutions to the field equations exist for
these symmetries and asymptotic behavior.

\subsection{Matter Solutions}

Consider next the case of Lifshitz solutions in the presence of a massive
gauge field $A^{\mu }$. The metric (\ref{met2}) and the gauge field (\ref
{gauge}) with $h(r)=1$ solve the field equation (\ref{E1}) provided
\begin{equation}
m^{2}=\frac{(n-1)z}{l^{2}}.  \label{mc}
\end{equation}

In order to have asymptotically Lifshitz solutions in 3rd-order Lovelock
gravity in the presence of matter, the following constraint
\begin{eqnarray}
q^{2} &=&\frac{2(z-1)L^{4}}{zl^{4}},  \notag \\
\Lambda  &=&-\frac{[(z-1)^2+n(z-2)+n^2]L^4+n(n-1)(\hat{\alpha}_{2}l^2-2\hat{\alpha}_{3})}{2l^{6}} \label{Coef5} \label{Coef5}
\end{eqnarray}
must hold for the cosmological constant and charge where we define
\begin{equation}
L^{4}\equiv l^{4}-2l^{2}\hat{\alpha}_{2}+3\hat{\alpha}_{3} \label{L4}
\end{equation}
for simplicity and we use this definition throughout the paper. Since $q^2>0$ we obtain the constraint
\begin{equation}
\frac{\hat{\alpha}_{2}}{l^{2}} < \frac{1}{2}+3\frac{\hat{\alpha}_{3}}{2 l^{4}} \label{a2}
\end{equation}
which in turn yields
\begin{equation}
 -\frac{(z-1)(n+z-1)(l^4+3 \hat{\alpha}_3)+n(n-1)(l^4+ \hat{\alpha}_3)}{2l^6}\leq
\Lambda< - \frac{n(n-1)}{4 l^2}\left(1- \frac{\hat{\alpha}_{3}}{ l^{4}}\right)  \label{LamCons}
\end{equation}
provided $\hat{\alpha}_{2} \geq 0$. Equation (\ref{LamCons}) shows that for $\hat{\alpha}_{3}<l^4$ the cosmological constant
is negative. For $\hat{\alpha}_{3}>l^4$, the cosmological constant can be positive
provided
\[
\frac{(z-1)(n+z-1)(l^4+3 \hat{\alpha}_3)+n(n-1)(l^4+ \hat{\alpha}_3)}{[2(z-1)(z+n-1)+n(n-1)]l^4}
< \frac{\hat{\alpha}_{2}}{l^{2}}  < \frac{1}{2}+3\frac{\hat{\alpha}_{3}}{2 l^{4}},
\]
where the last inequality comes from the condition (\ref{a2}).

Note that for
Einstein gravity ($\hat{\alpha}_{2}=\hat{\alpha}_{3}=0$) $L=l$ and the above
conditions reduce to those which are given in \cite{Peet} for $n=3$. In
Gauss-Bonnet gravity ($\alpha _{3}=0$), these conditions become
\begin{eqnarray}
q^{2} &=&\frac{2(z-1)(l^{2}-2\hat{\alpha}_{2})}{zl^{2}},  \notag \\
\Lambda  &=&-\frac{[(z-1)^2+n(z-2)+n^2](l^{2}-2\hat{\alpha}_{2})+n(n-1)\hat{\alpha}_{2}}{2l^{4}}.  \label{Coef4}
\end{eqnarray}
where now $l^{2}$ must be larger than $2\hat{\alpha}_{2}$. If $\hat{\alpha}%
_{2}=l^{2}/2$, then the charge $q$ becomes zero and the conditions (\ref
{Coef4}) reduce to the conditions (\ref{Coef1}) as expected. More generally,
if $L=0$, then $q$ vanishes and the conditions (\ref{Coef5}) reduce to the
conditions (\ref{Coef2}) as expected.

In the absence of a cosmological constant, the Lifshitz solution exists
provided
\begin{eqnarray}
q^{2} &=&\frac{2n(n-1)(z-1)(\hat{\alpha}_{2}-2l^{2})}{%
zl^{2}[3(z-1)^{2}+3nz+n(n-4)]},  \notag \\
\hat{\alpha}_{3} &=&\frac{(2\hat{\alpha}_{2}-l^{2})\left[ (z-1)^{2}+nz\right]
+n\left[ (n-3)\hat{\alpha}_{2}-(n-2)l^{2}\right] }{[3(z-1)^{2}+3nz+n(n-4)]}l^2. \label{Coef6}
\end{eqnarray}
In this case the conditions (\ref{Coef6}) reduce to the condition (\ref
{Coef3}) if $\hat{\alpha}_{2}=2l^{2}$, leading to $q=0$ and the absence of
matter.

\section{Asymptotic Lifshitz Black holes \label{black}}

In this section we seek black hole solutions in 3rd-order Lovelock gravity
that are asymptotically Lifshitz. We therefore consider the field equations (%
\ref{E1})-(\ref{E3}) with the conditions (\ref{Coef5}), but with the more
general ansatz (\ref{met1}) and (\ref{gauge}).

\subsection{Series solutions near the horizon}
We now consider the near-horizon behavior of such solutions. Requiring that
$f(r)$ and $g(r)$ go to zero linearly, that is
\begin{eqnarray*}
f(r)
&=&f_{1}\left\{(r-r_{0})+f_{2}(r-r_{0})^{2}+f_{3}(r-r_{0})^{3}+f_{4}(r-r_{0})^{4}+...\right\},
\\
g(r)
&=&g_{1}(r-r_{0})+g_{2}(r-r_{0})^{2}+g_{3}(r-r_{0})^{3}+g_{4}(r-r_{0})^{4}+...,
\\
h(r)
&=&f_{1}^{1/2}\left\{h_{0}+h_{1}(r-r_{0})+h_{2}(r-r_{0})^{2}+h_{3}(r-r_{0})^{3}+h_{4}(r-r_{0})^{4}+...\right\},
\end{eqnarray*}
and inserting these expansions into the equations of motion arising from
Eqs. (\ref{E1})-(\ref{E2}) with the conditions (\ref{Coef5}), and solving
for the various coefficients, we find that $h_{0}=0$. This is consistent
with the fact that the flux $dA$ should go to a constant at the horizon.
Also, one may note that by scaling time we can adjust the constant $f_{1}$\
by an overall multiplicative factor (note the use of $f_{1}^{1/2}$\ in the
expansion of $h(r)$\ as well, which is due to $dt$\ in the one-form $A$).
We find the following constraint on the $1$st order constants
\begin{eqnarray}
g_{1} &=&\frac{z}{r_{0}^{3}}\Big\{3\hat{\alpha}_{3}\left[ {h_{1}}^{2}(z-1){%
r_{0}}^{5}+{k}^{2}z(n-1){l}^{4}\right] \notag \\
&&+2{r_{0}}^{2}l^{2}\hat{\alpha}_{2}\left[ -{h_{1}}^{2}(z-1){r_{0}}^{3}+{l}%
^{2}kz(n-1)\right] +{r_{0}}^{4}{l}^{4}\left[ {h_{1}}^{2}\left( z-1\right)
r_{0}+z\left( n-1\right) \right] \Big\}^{-1} \notag\\
&&\times \Big\{\left\{ \left[ 3(z-1)^{2}+3nz++n(n-4)\right] {r_{0}}^{6}+{k}%
(n-1)(n-6){l}^{6}\right\} \hat{\alpha}_{3} \notag\\
&&-{{r_{0}}}^{2}{l}^{2}\hat{\alpha}_{2}\left\{ \left[ 2(z-1)^{2}+2nz+n(n-3)%
\right] {r_{0}}^{4}+{k}^{2}{l}^{4}(n-1)(n-4)\right\} \notag \\
&&+{r_{0}}^{4}{l}^{4}\left[ (z-1)^{2}+nz+n(n-1)\right] {r_{0}}%
^{2}+(n-1)(n-2)k{l}^{2}\Big\},
\end{eqnarray}
which means that not all boundary conditions are allowed. Note that here the
coefficient $h_{1}$ is arbitrary, and should be chosen suitable in order to
explore the numerical solutions. The higher order coefficients may be found
easily, but their expressions are very lengthy and so we don't write them here for  reasons of economy.

One may also investigate the behavior of the metric functions at large $r$.
We relegate this to the Appendix, where we show that the powers of $1/r$ for $k=0$ are in general
non-integer.
\subsection{Numeric Solutions}\label{numsol}
In order to find the numeric black hole solutions of the system given by
Eqs. (\ref{E1})-(\ref{E3}) with the conditions (\ref{Coef5}), we define
\begin{equation}
\frac{dh}{dr}\equiv j(r),  \label{dhr}
\end{equation}
and find the first derivatives of $j(r)$, $f(r)$ and $g(r)$ as
\begin{equation}
\frac{dj}{dr} =\frac{(n-1)zh-\left[ z(n+z-2)h+(n+2z-1)rj\right] g}{r^{2}g}+%
\frac{(z-1)L^{4}(zh+rj)r^{4}h^{2}}{r^{2}fgH},  \label{djr}
\end{equation}
\begin{eqnarray}
\frac{df}{dr} &=&\frac{1}{(n-1)zr^{3}gH}\Big\{-(n-1)[n+6(z-1)]z\hat{\alpha}%
_{3}r^{6}fg^{3} \notag\\
&& +(n-1)zl^{2}r^{4}\left[ 3k\hat{\alpha}_{3}(n+4z-6)+(n+4z-4)%
\hat{\alpha}_{2}r^{2}\right] fg^{2}  \notag \\
&&-(n-1)zl^{4}r^{2}\left[ 3(n+2z-6)k^{2}\hat{\alpha}_{3}+2(n+2z-4)k\hat{%
\alpha}_{2}r^{2}+(n+2z-2)r^{4}\right] fg  \notag \\
&&+z\hat{\alpha}_{3}\left\{[3(z-1)^{2}+3nz+n(n-4)]r^{6}+(n-1)(n-6)kl^{6}\right\}f  \notag \\
&&-z\hat{\alpha}_{2}l^{2}r^{2}\left\{
[2(z-1)^{2}+2nz+n(n-3)]r^{4}-(n-1)(n-4)k^{2}l^{4}\right\} f  \notag \\
&&+zl^{4}r^{4}\left\{ [(z-1)^{2}+nz+n(n-1)]r^{2}+(n-1)(n-2)kl^{2}\right\} f
\notag \\
&&-(z-1)L^{4}r^{6}[(zh+rj)^{2}g+(n-1)zh^{2}]\Big\},
\end{eqnarray}
\begin{eqnarray}
\frac{dg}{dr} &=&\frac{1}{(n-1)zr^{3}fH}\Big\{-n(n-1)z\hat{\alpha}%
_{3}r^{6}fg^{3}+(n-1)zl^{2}r^{4}\left[ 3k\hat{\alpha}_{3}(n-2)+n\hat{\alpha}%
_{2}r^{2}\right] fg^{2}  \notag \\
&&-(n-1)zl^{4}r^{2}\left[ 3(n-4)k^{2}\hat{\alpha}_{3}+2(n-2)k\hat{\alpha}%
_{2}r^{2}+nr^{4}\right] fg  \notag \\
&&+z\hat{\alpha}_{3}\left\{[3(z-1)^{2}+3nz+n(n-4)]r^{6}+(n-1)(n-6)kl^{6}\right\}f  \notag \\
&&-z\hat{\alpha}_{2}l^{2}r^{2}\left\{
[2(z-1)^{2}+2nz+n(n-3)]r^{4}-(n-1)(n-4)k^{2}l^{4}\right\} f  \notag \\
&&+zl^{2}r^{2}\left\{ [(z-1)^{2}+nz+n(n-1)]r^{2}+(n-1)(n-2)kl^{2}\right\} f
\notag \\
&&-(z-1)L^{4}r^{6}[(zh+rj)^{2}g+(n-1)zh^{2}]\Big\}, \label{dgr}
\end{eqnarray}
where $H$ is
\begin{equation*}
H=3\hat{\alpha}_{3}(kl^{2}-r^{2}g)^{2}+\hat{\alpha}_2%
l^{2}r^{2}(kl^{2}-r^{2}g)+l^{4}r^{4}.
\end{equation*}
Now having a system of 4 first order ordinary differential equations, one
may explore the numeric solutions by choosing suitable initial conditions
for $f(r_{0})$, $g(r_{0})$, $h(r_{0})$ and $j(r_{0})$, where $r_{0}$ is the
radius of horizon. This can be done by using the method explained in \cite
{Mann}.
\begin{figure}[h]
\epsfxsize=10cm \centerline{\epsffile{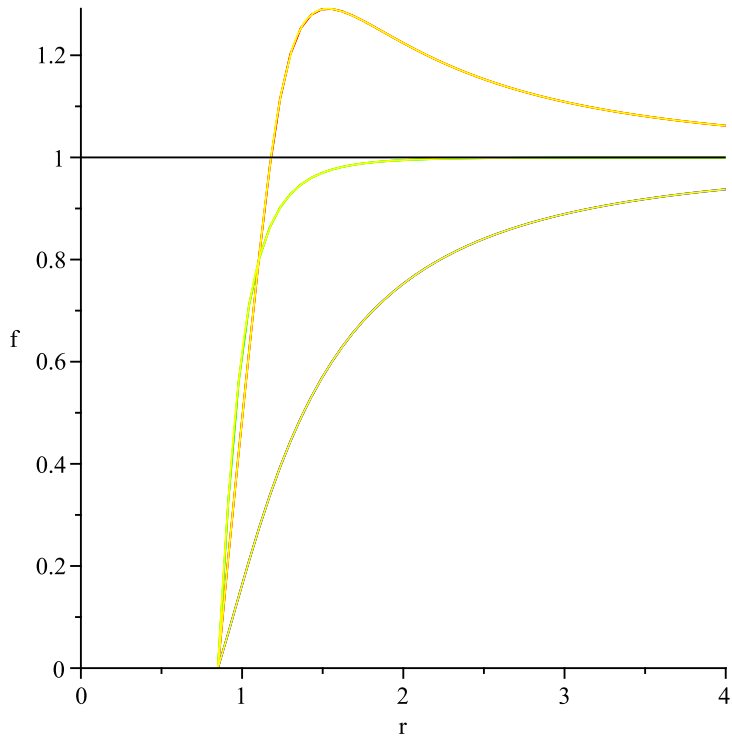}}
\caption{$f(r)$, $g(r)$ and exact solution versus $r$ for $n=6$, $l=1$, $z=1$, $\hat{\protect\alpha}_{2}=.25$, $\hat{\protect\alpha}%
_{3}=.3$ and $r_{0}=0.85$ for $k=1$ (top),
$k=0$ (middle) and $k=-1$ (bottom). Note that $f(r)$, $g(r)$ and the exact solution
lie on each other for each $k$.}
\label{Fz1k}
\end{figure}
First, we apply the numerical method for $z=1$ which can be solved
exactly, and compare the numerical solution to the exact one.
Indeed, for $z=1$, the charge vanishes, and the problem reduces to
the case of black holes in third order Lovelock gravity with specific
choices of Lovelock coefficients. In this case,  we obtain the exact
solutions of Lovelock gravity \cite{Deh}, which have  $f(r)=g(r)$.  The diagrams of
the functions $f(r)$ and $g(r)$ versus $r$
for $z=1$ with $k=0$ and $k=\pm1$ have been shown in Fig \ref{Fz1k}. In this figure
the exact solutions have also been shown. As one can see from the plots, the
numerically obtained functions $f(r)$
and $g(r)$  are equal and both of them are exactly lie on the exact solution within limits of numerical precision.
We regard this as a good test of our methods.
Note that for $k=0,-1$ and $z=1$ (the solution without matter) , the metric
functions increase from zero at $r=r_{0}$ to $1$ at $r=\infty $. However for $k=1$ and
$z=1$, the metric functions grow to
be larger than unity at intermediate values of $r$ and
asymptote to 1 at infinity.
\begin{figure}[h]
\textit{\epsfxsize=10cm \centerline{\epsffile{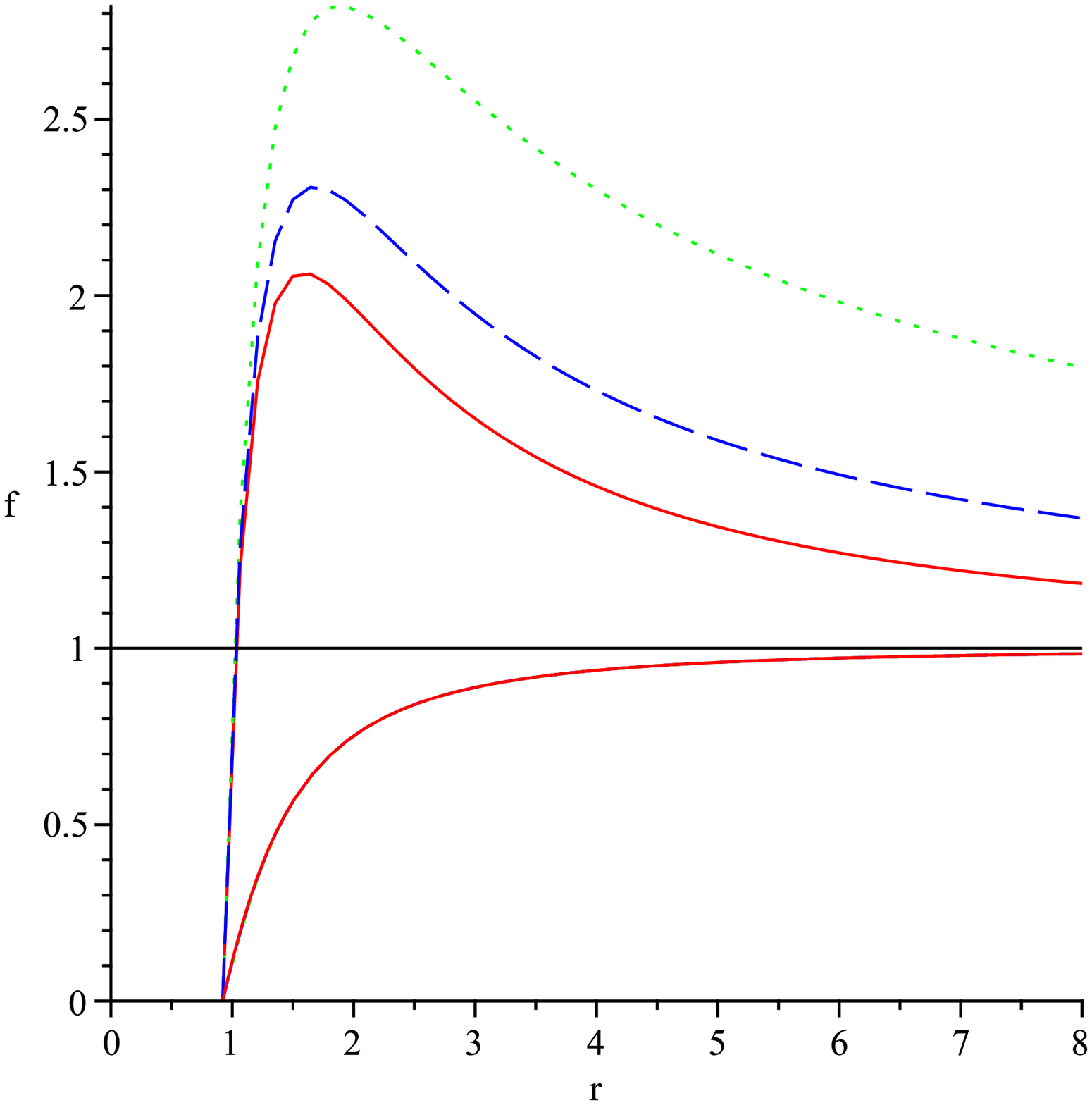}} }
\caption{$f(r)$ versus $r$ for $k=-1$%
, $n=6$, $l=1$,  $\hat{\protect\alpha}_{2}=.25$, $\hat{\protect\alpha}%
_{3}=.3$, $r_{0}=0.92$ in 3rd order Lovelock,
Gauss-Bonnet and Einstein gravities red (solid), green (dotted) and blue (dashed), respectively.
The three upper curves are $f(r)$ for $z=2$ and the three lower curves,
which are nearly identical and so lie on top of each other, correspond to $z=1$.}
\label{Fskm1}
\end{figure}
\begin{figure}[h]
\textit{\epsfxsize=10cm \centerline{\epsffile{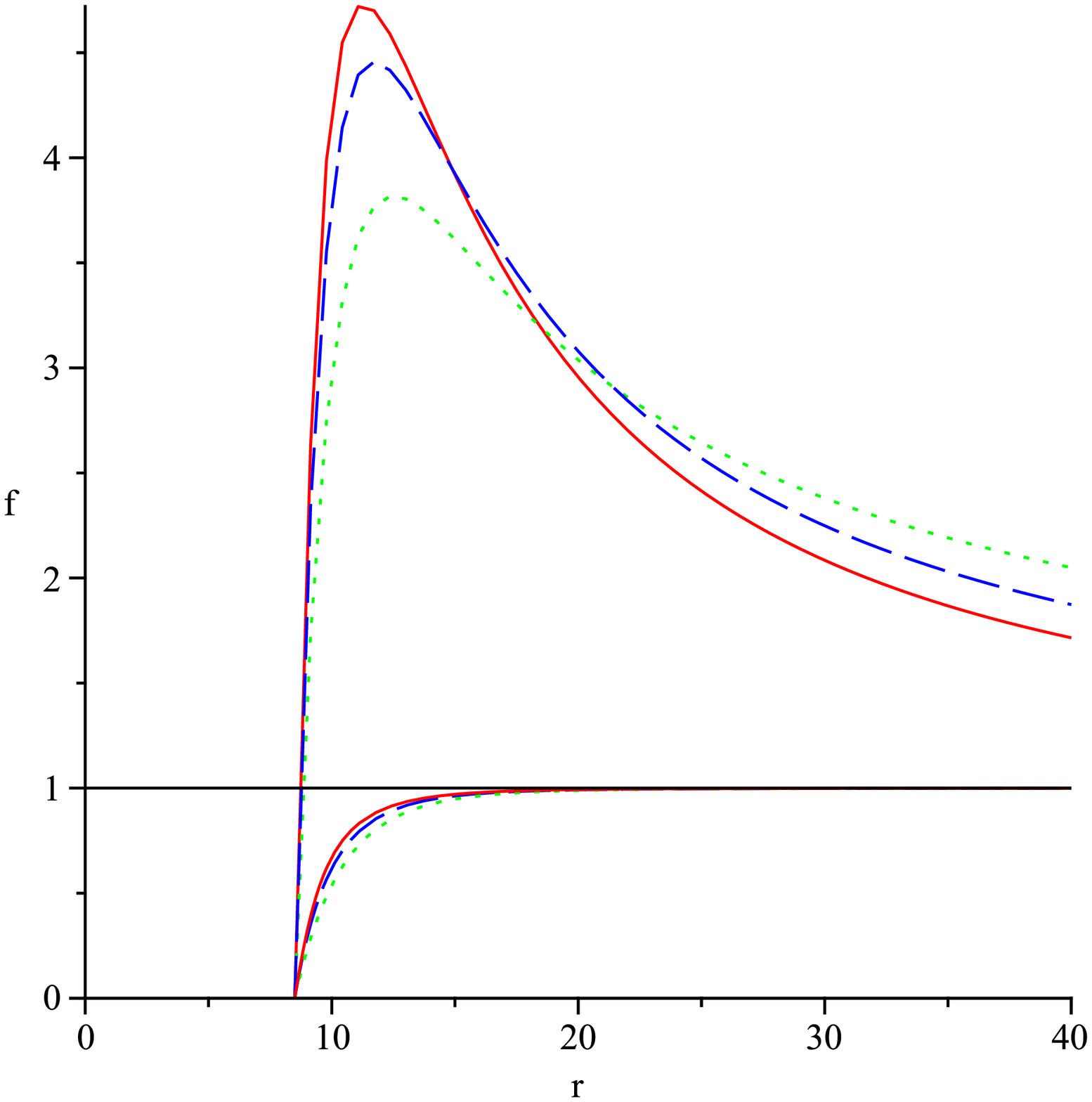}} }
\caption{$f(r)$ versus $r$ for $k=-1$%
, $n=6$, $l=1$,  $\hat{\protect\alpha}_{2}=.25$, $\hat{\protect\alpha}%
_{3}=.3$, $r_{0}=8.5$ in 3rd order Lovelock,
Gauss-Bonnet and Einstein gravities red (solid), green (dotted) and blue (dashed), respectively.
The three upper curves are $f(r)$ for $z=2$ and the three lower curves
correspond to $z=1$.}
\label{Flkm1}
\end{figure}
\begin{figure}[h]
\textit{\epsfxsize=10cm \centerline{\epsffile{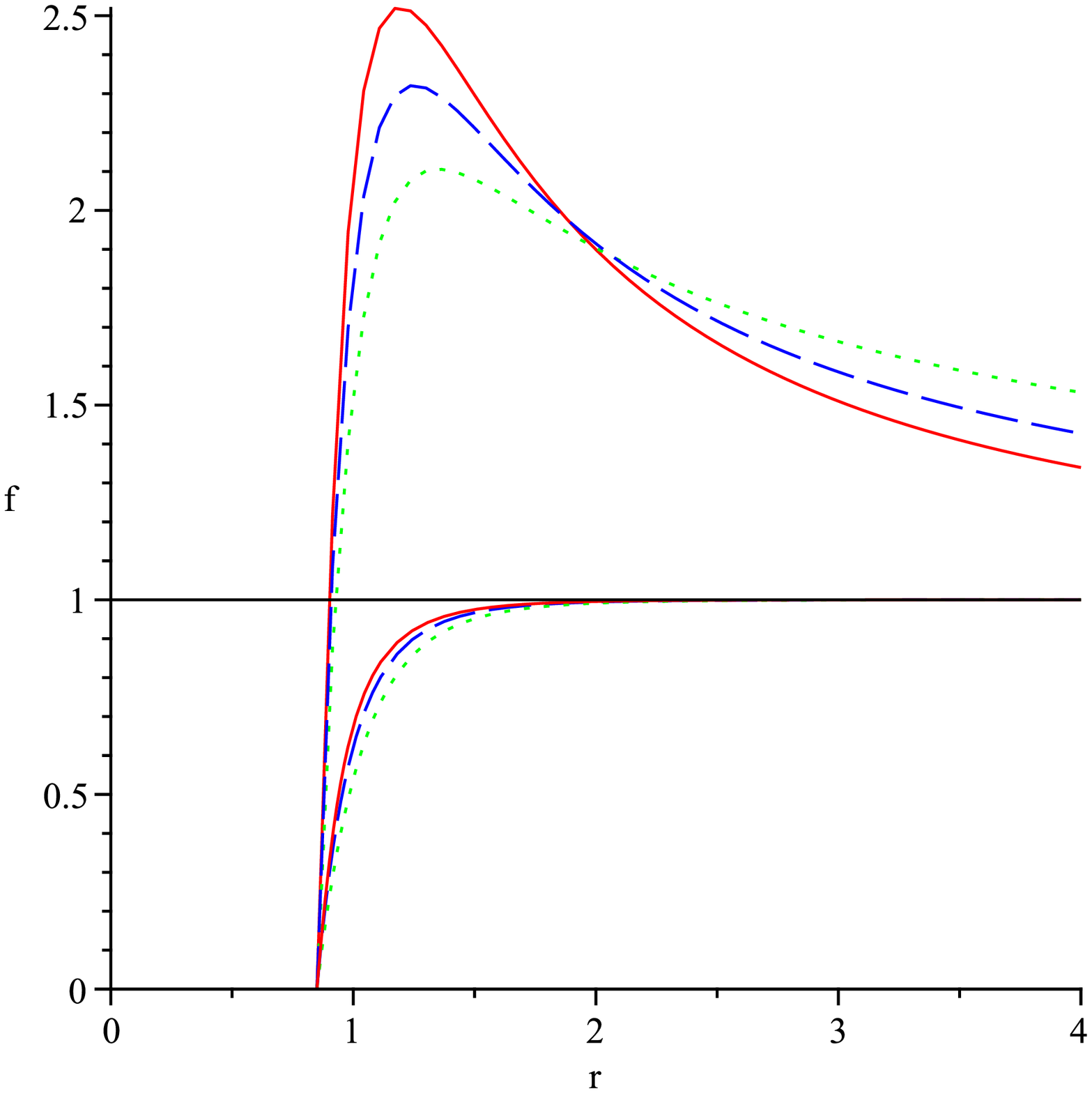}} }
\caption{$f(r)$ versus $r$ for $k=0$%
, $n=6$, $l=1$,  $\hat{\protect\alpha}_{2}=.25$, $\hat{\protect\alpha}%
_{3}=.3$, $r_{0}=0.85$ in 3rd order Lovelock,
Gauss-Bonnet and Einstein gravities red (solid), green (dotted) and blue (dashed), respectively.
The three upper curves are $f(r)$ for $z=2$ and the three lower curves
correspond to $z=1$.}
\label{Fsk0}
\end{figure}
\begin{figure}[h]
\textit{\epsfxsize=10cm \centerline{\epsffile{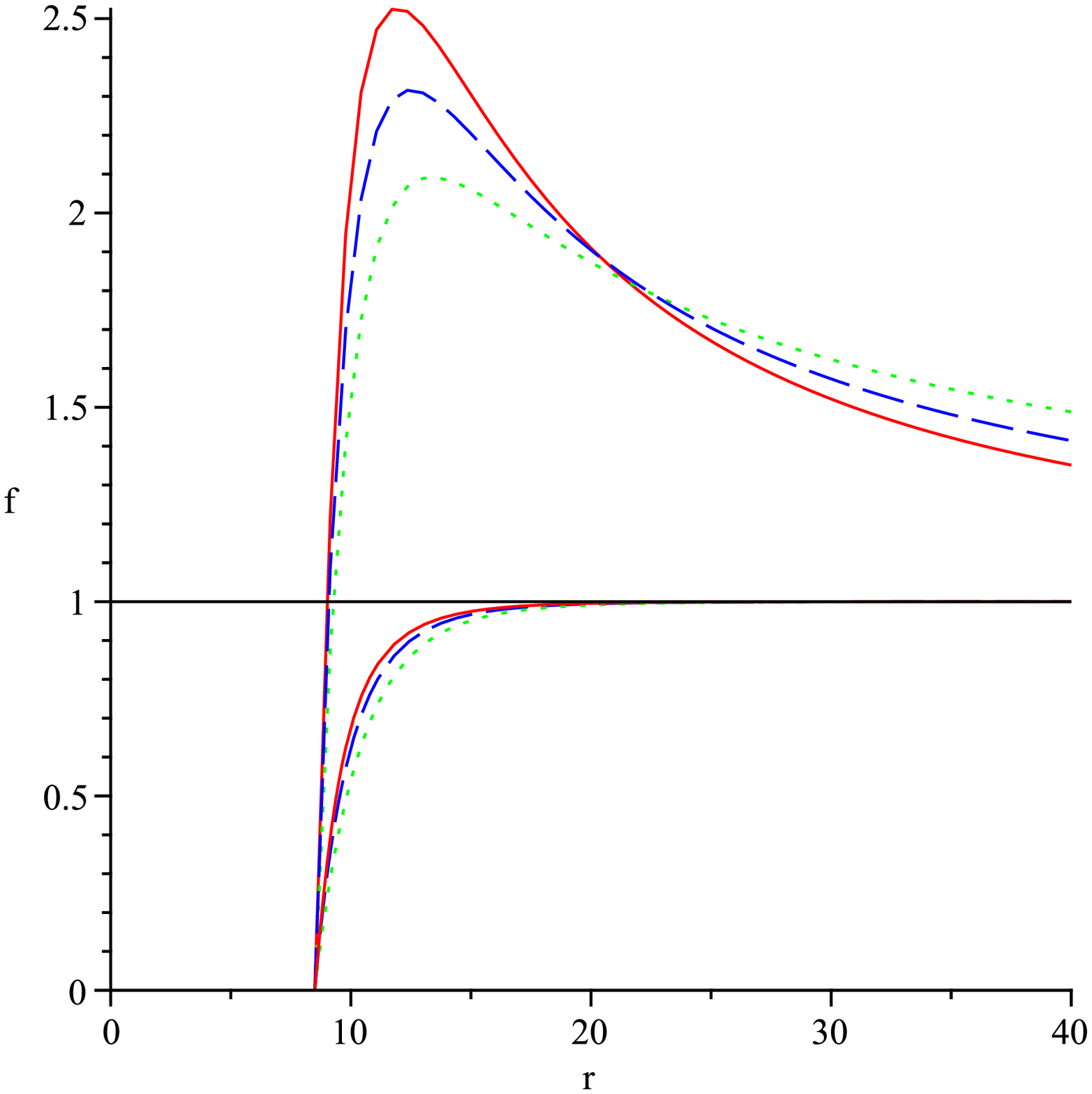}} }
\caption{$f(r)$ versus $r$ for $k=0$%
, $n=6$, $l=1$,  $\hat{\protect\alpha}_{2}=.25$, $\hat{\protect\alpha}%
_{3}=.3$, $r_{0}=8.5$ in 3rd order Lovelock,
Gauss-Bonnet and Einstein gravities red (solid), green (dotted) and blue (dashed), respectively.
The three upper curves are $f(r)$ for $z=2$ and the three lower curves
correspond to $z=1$.}
\label{Flk0}
\end{figure}
\begin{figure}[tbp]
\textit{\epsfxsize=10cm \centerline{\epsffile{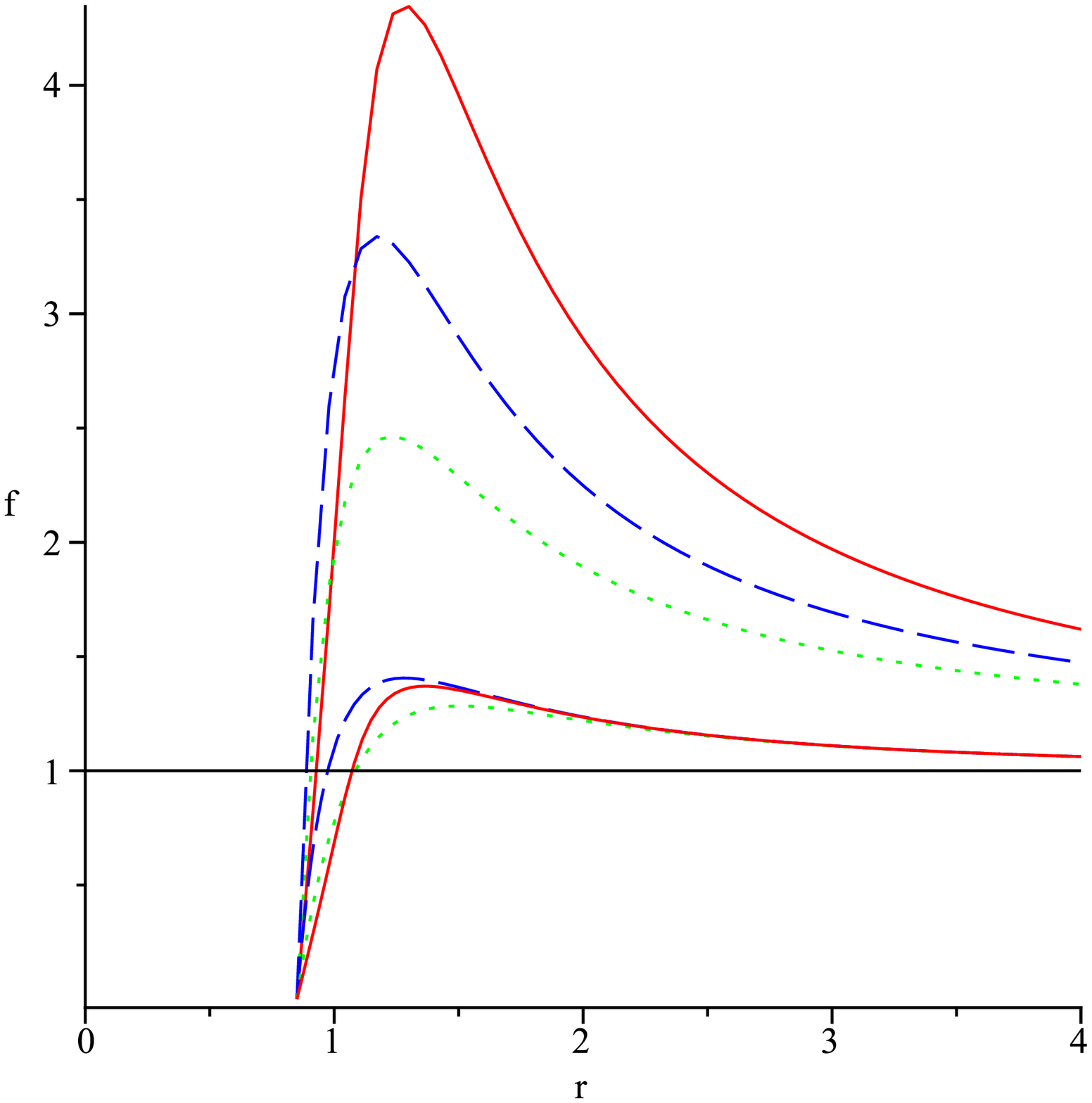}} }
\caption{$f(r)$ versus $r$ for $k=1$%
, $n=6$, $l=1$,  $\hat{\protect\alpha}_{2}=.25$, $\hat{\protect\alpha}%
_{3}=.3$, $r_{0}=0.85$ in 3rd order Lovelock,
Gauss-Bonnet and Einstein gravities red (solid), green (dotted) and blue (dashed), respectively.
The three upper curves are $f(r)$ for $z=2$ and the three lower curves
correspond to $z=1$.}
\label{Fsk1}
\end{figure}
\begin{figure}[tbp]
\textit{\epsfxsize=10cm \centerline{\epsffile{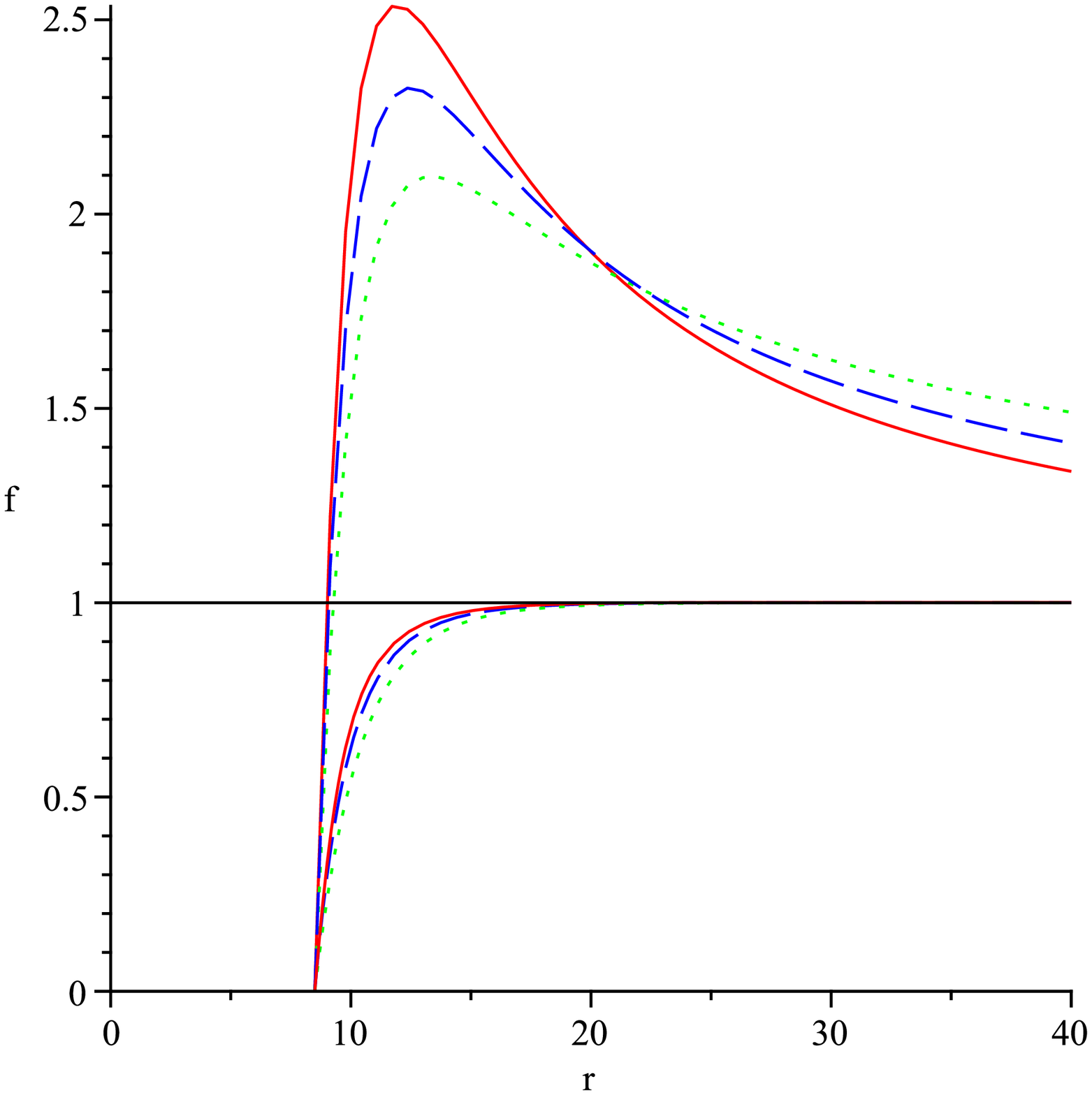}} }
\caption{$f(r)$ versus $r$ for $k=1$%
, $n=6$, $l=1$,  $\hat{\protect\alpha}_{2}=.25$, $\hat{\protect\alpha}%
_{3}=.3$, $r_{0}=8.5$ in 3rd order Lovelock,
Gauss-Bonnet and Einstein gravities red (solid), green (dotted) and blue (dashed), respectively.
The three upper curves are $f(r)$ for $z=2$ and the three lower curves
correspond to $z=1$. }
\label{Flk1}
\end{figure}
Now, we consider the solutions for $z\neq 1$, for which  matter
exists, where we must solve the system of equations numerically.
Figure \ref{Fskm1} shows the function $f(r)$ as a function of $r$ for a small black
hole of Einstein, Gauss-Bonnet and Lovelock gravity with $z=1$ and $z=2$. As one can see from this figure,
while $f(r)$ for Einstein, Gauss-Bonnet and Lovelock with  $z=1$ (solution without matter) are almost the
same, they are different for $z=2$  for the same values of  $\alpha_2$ and $\alpha_3$.
Furthermore, for $z=2$,
the metric function $f(r)$ rapidly increases to values larger than 1,
eventually asymptoting to $1$ as $r\rightarrow \infty $, while for $z=1$
this function monotonically increases from zero at $r=r_{0}$ to $1$ at $r=\infty $. Thus, it appears that
asymptotic Lifshitz solutions are more sensitive than are their $z=1$ (AdS) counterparts
to the corrections induced by Lovelock gravity. This feature
also occurs for large black holes with $k=-1$ (see Fig. \ref{Flkm1}).
Note that for $k=-1$, the function $f(r)$ in Gauss-Bonnet gravity is larger than
that in third order Lovelock gravity. This is due to the fact that
the slope of the function $f(r)$ decreases as $\hat{\alpha}_3$ increases (see the terms
in $df/dr$ with the factor $\hat{\alpha}_3$ in Eq. (\ref{dgr}) with $k=-1$, which are negative).
Figures \ref{Fsk0}-\ref{Flk0} show that this fact also occurs for the small and large black holes with $k=0$.
For the case of $k=1$, one can see from Figs \ref{Fsk1} and \ref{Flk1} that both the asymptotic
AdS and Lifshitz black holes are different for various order of Lovelock
gravities, although the difference between the metric function $f(r)$ for various
order of Lovelock gravity becomes more relevant as $z$ increases.

\begin{figure}[tbp]
\textit{\epsfxsize=10cm \centerline{\epsffile{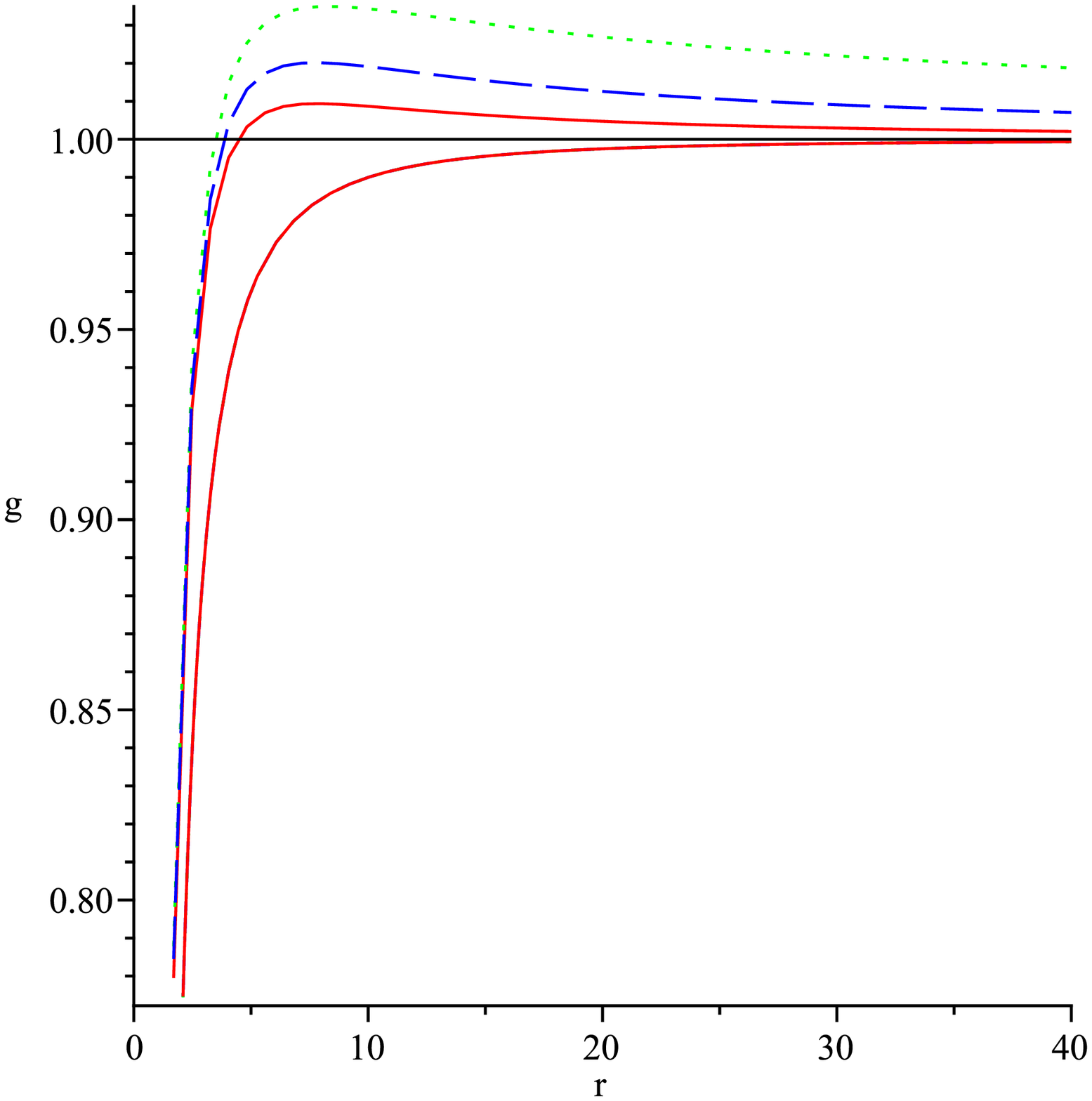}} }
\caption{$g(r)$ versus $r$ for $k=-1$%
, $n=6$, $l=1$,  $\hat{\protect\alpha}_{2}=.25$, $\hat{\protect\alpha}%
_{3}=.3$, $r_{0}=0.85$ in 3rd order Lovelock,
Gauss-Bonnet and Einstein gravities red (solid), green (dotted) and blue (dashed), respectively.
The three upper curves are $f(r)$ for $z=2$ and the three lower curves,
which lie on each other, correspond to $z=1$.}
\label{gskm1}
\end{figure}

We find that the dependence on $z$ of $g(r)$
is not as significant as for $f(r)$. For this and reasons of economy, we plot only
the function $g(r)$ in one case (see Fig. \ref{gskm1}). We also find that
 $g(r)$ is monotonically increasing for topological black holes with $k=0$, whereas (as noted previously)
 $f(r)$ has a maximum larger than $1$ at $r$ several times $r_0$.

We pause to comment on the relative importance of the various terms in the Lovelock action.
While there is no essential problem in considering all terms to be similar in magnitude, we obtain
some insight into Lovelock gravity by treating it as an approximation (in powers of the curvature)
to a full quantum theory of gravity. For simplicity we shall consider  solutions
obtained in this paper for  $k=0$, which asymptote to the Lifshitz background (\ref{met2}). In this case the $p$th order Lagrangian of Lovelock gravity for the metric (\ref{met1}) may be written as:
\begin{equation}
\mathcal{L}'_p=(-1)^{p-1}\frac{(n-1)}{(2p-1)l^{z+2p-1}r^{(2p-1)(z-1)}}\left(\frac{g}{f}\right)^{p-1/2}
\frac{d}{dr}\left\{r^{n+2p(z-1)}f^p\right\}, \label{Lagp}
\end{equation}
where   $\mathcal{L}'_p=(\alpha_p/\hat{\alpha}_p)\sqrt{-g}\mathcal{L}_p$. The above expression for the Lagrangian is correct for any value of $p$, though in this paper we have considered only
$p\leq 3$. The Lagrangian (\ref{Lagp}) can be evaluated on the solutions at large $r$ by the use of the expansions of the metric functions  which are given in the Appendix. It is a matter of calculation to show that the ratio $\mathcal{L}'_{p+1}/\mathcal{L}'_{p}$ at large $r$ is
\begin{equation}
\frac{\mathcal{L}'_{p+1}}{\mathcal{L}'_{p}}=-\frac{(2p-1)[n+2(p+1)(z-1)]}{(2p+1)[n+2p(z-1)]l^2},
\end{equation}
and so ${\mathcal{L}'_{p+1}}<{\mathcal{L}'_{p}}/l^2$.

\section{Thermodynamics of black holes \label{Therm}}

The entropy of a black hole in Lovelock gravity is \cite{Wald}
\begin{equation}
S=\frac{1}{4}\sum_{k=1}^{p}k\alpha _{k}\int d^{n-1}x\sqrt{\tilde{g}}\tilde{%
\mathcal{L}}_{k-1},  \label{Enta}
\end{equation}
where the integration is done on the $(n-1)$-dimensional spacelike
hypersurface of the Killing horizon with induced metric $\tilde{g}_{\mu \nu }$
(whose determinant is $\tilde{g}$),  and $%
\tilde{\mathcal{L}}_{k}$ is the $k$th order Lovelock Lagrangian of $\tilde{g}%
_{\mu \nu }$. It is a matter of calculation to show that the entropy
 of a black hole per unit volume of the horizon in third order Lovelock
gravity is
\begin{equation}
S=\frac{r_{0}^{n-1}}{4}\left( 1+\frac{2k(n-1)\hat{\alpha}_{2}}{(n-3)r_{0}^{2}%
}+\frac{3k^{2}(n-1)\hat{\alpha}_{3}}{(n-5)r_{0}^{4}}\right).  \label{Ent}
\end{equation}
This reduces to the area law of entropy for $\hat{\alpha}_{2}=\hat{\alpha}%
_{3}=0$.

One can obtain the temperature of the event horizon by using standard
Wick-rotation methods, yielding the result
\begin{equation}
T=\left(\frac{r^{z+1}\sqrt{f^{\prime }g^{\prime }}}{4 \pi l^{z+1}}\right)_{r=r_0}.
\end{equation}

For the case of asymptotic Lifshitz black hole with (identified) hyperbolic horizon, the
temperature of the exact solution (\ref{exa}) with horizon radius $r_{0}=l$ in vacuum
is
\begin{equation}
T=\frac{1}{2\pi l}.
\end{equation}

\begin{figure}[tbp]
\textit{\epsfxsize=10cm \centerline{\epsffile{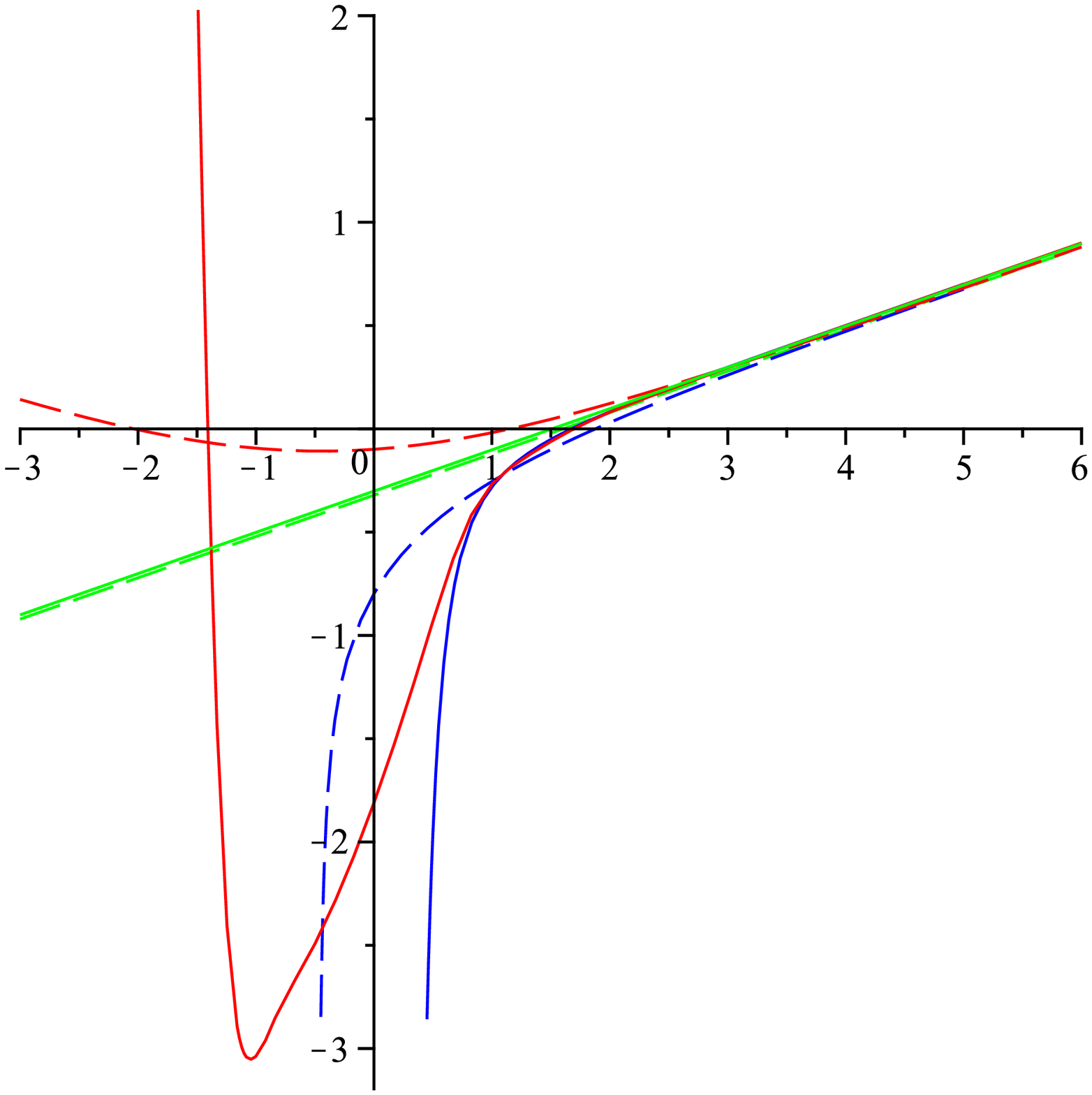}} }
\caption{$\log T$ versus $\log S$ for $n=6$, $l=1$, $z=1$ in Einstein
and Lovelock ($\hat{\protect\alpha}_{2}=.25$, $\hat{\protect\alpha}%
_{3}=.3$) gravities with $k=1$ (red), $k=0$ (green) and $k=-1$
(blue), where the dashed and solid lines are curves in Einstein and Lovelock gravities, respectively.}
\label{TKz1}
\end{figure}

\begin{figure}[tbp]
\textit{\epsfxsize=10cm \centerline{\epsffile{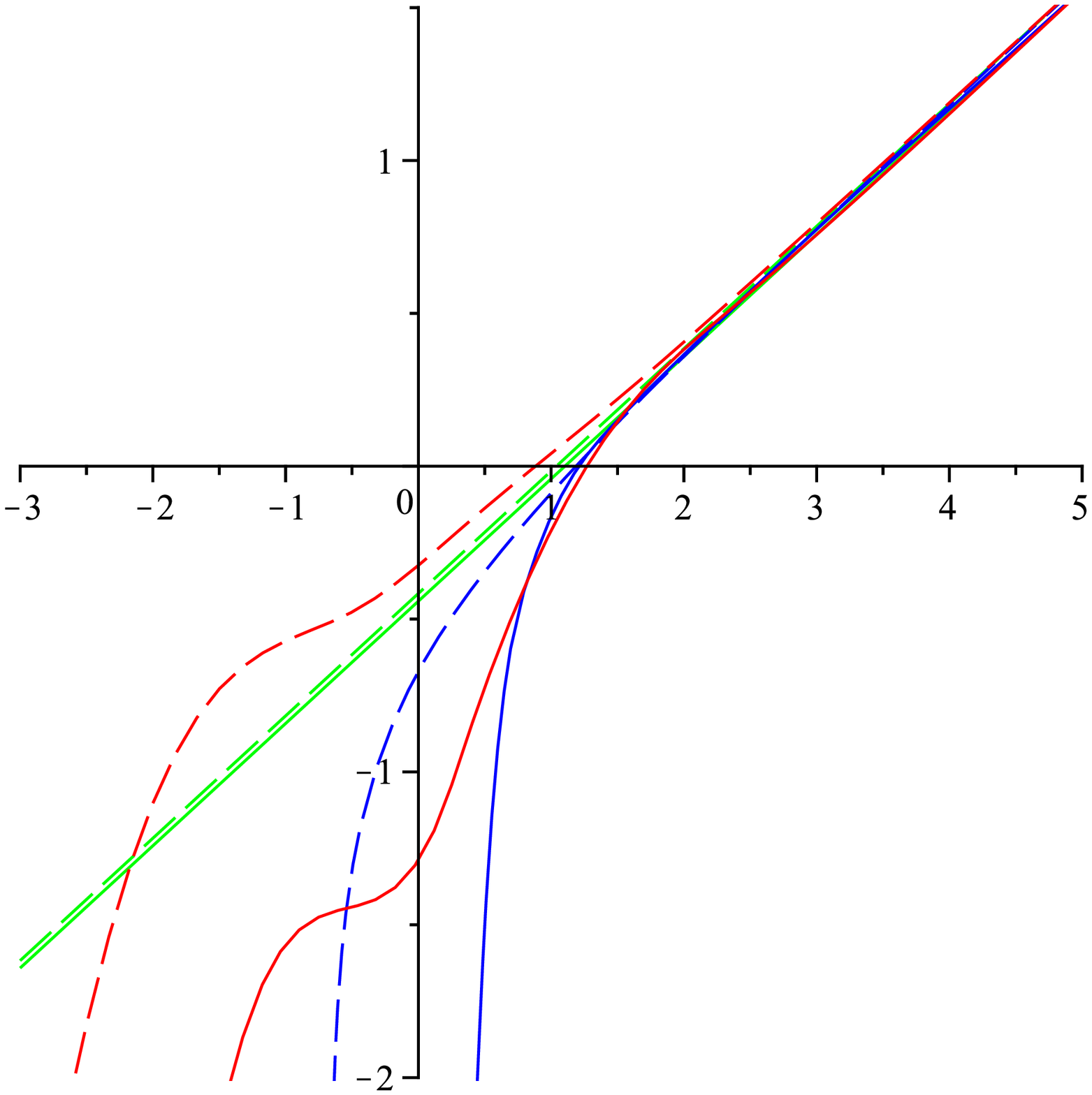}} }
\caption{$\log T$ versus $\log S$ for $n=6$, $l=1$, $z=2$ in Einstein
and Lovelock ($\hat{\protect\alpha}_{2}=.25$, $\hat{\protect\alpha}%
_{3}=.3$) gravities with $k=1$ (red), $k=0$ (green) and $k=-1$
(blue), where the dashed and solid lines are curves in Einstein and Lovelock gravities, respectively.}
\label{TKz2}
\end{figure}

The temperature of more general Lovelock-Lifshitz black holes can be
calculated numerically. First, we review the thermodynamics of asymptotic AdS
black holes ($z=1$) to compare it with the asymptotic Lifshitz black holes. As one can see from figure \ref{TKz1},
the logarithm of the temperature of black holes versus the logarithm of the entropy
for large black holes is linear, while for small black holes this occurs only
in the case of $k=0$. For $k=-1$, one encounters with an extreme black hole in Einstein (blue dashed line)
and Lovelock gravity (blue solid line). However, the horizon radius of the extreme black
hole of Lovelock gravity is smaller than that of Einstein gravity. Note that
the slopes of the graphs of $\log T$ versus $\log S$ for small black holes of Einstein and
Lovelock gravity with $k=1$ are negative, and therefore they are unstable. The unstable phase in Lovelock gravity occurs for a smaller radius black hole than in Einstein theory.
Plotting $\log T$ versus $\log S$ for the
case of asymptotic Lifshitz black holes with $z=2$, we see from Fig. \ref{TKz2} that
the temperature of black holes in Lovelock gravity with given entropy $S$ is smaller than
the temperature of black holes of Einstein gravity with the same entropy. Conversely, at a given
temperature the entropy of the Lovelock black holes -- extremal and non-extremal -- is larger than for the Einstein ones for both $z=1$ and $z=2$.
The horizon radii of  extreme black holes in both Einstein and Lovelock gravity for $k=-1$ with $z=2$ are smaller than their  $z=1$ counterparts, and
for a given $z$ smaller in Lovelock gravity than in Einstein gravity.

Numerical calculations show that there is no unstable phase for $z=2$ black holes for $k=1$.
We also find that for $k=0$ and the large black
holes for $k=\pm 1$, the temperature is proportional to $r_0^z$, while the entropy is proportional
to $r_0^{n-1}$, yielding
\begin{equation}
T \varpropto S^{z/(n-1)}.
\end{equation}
for the dependence of $T$ on $S$.
In order to say more about the temperature, one should compute the
energy density of these black holes using (for example) the counterterm method.  While some work has been done on this for asymptotic Lifshitz solutions of Einstein gravity \cite{Ros,Peet2}, the corresponding formalism needs to be developed  for Lovelock gravity, an endeavour that we hope to address in the future.

\section{Rotating Lovelock-Lifshitz solutions}

\label{rot} In this section we endow the Lifshitz spacetime with a global
rotation. We first consider the Lifshitz solution with one rotation
parameter. We write the metric as
\begin{equation}
ds^{2}=-\frac{r^{2z}}{l^{2z}}dt^{2}+\frac{l^2 dr^{2}}{r^{2}}+r^{2}d\theta
_{1}^{2}+r^{2}\sum\limits_{i=2}^{n-1}d\theta _{i}^{2},  \label{metR1}
\end{equation}
In order to add angular momentum to the spacetime, we perform the following
rotation boost in the $t$-$\theta _{1}$ plane:
\begin{equation}
t\mapsto \Xi t-a\theta _{1}\ \ \ \ \ \ \ \ \ \ \theta _{1}\mapsto \Xi \theta
_{1}-\frac{a}{l^{2}}t,  \label{tth}
\end{equation}
where $a$ is the rotation parameter and $\Xi =1+a^{2}/l^{2}$. Substituting
eq. (\ref{tth}) into eq. (\ref{metR1}) we obtain
\begin{equation}
ds^{2}=-\frac{r^{z}}{l^{z}}\left( \Xi dt-ad\theta _{1}\right) ^{2}+\frac{%
l^2 dr^{2}}{r^{2}}+r^{2}\left( \frac{a}{l^{2}}dt-\Xi d\theta _{1}\right)
^{2}+r^{2}\sum\limits_{i=2}^{n-1}d\theta _{i}^{2}.  \label{metR2}
\end{equation}

The transformation (\ref{tth}) generates a new metric if $\theta_1$ is periodically
identified since the transformation (\ref{tth})   can be done
locally but not globally \cite{Stach}. The periodic
nature of $\theta _{1}$ allows the metrics (%
\ref{metR1}) and (\ref{metR2}) to be locally mapped into each other but not
globally, and so they are distinct. It is a matter of straightforward calculation to
show that the metric (\ref{metR2}) is a solution of Lovelock gravity,
provided one of the conditions (\ref{Coef2})-(\ref{Coef1}) holds.
Furthermore the metric (\ref{metR2}) is also a solution to the field equations in the
presence of the vector field
\begin{equation}
A=q\frac{r^{z}}{l^{z}}\left( \Xi dt-ad\theta _{1}\right) ,
\end{equation}
provided the condition (\ref{mc}) and one of the conditions (\ref{Coef4})-(%
\ref{Coef6}) are satisfied.

It is straightforward to generalize the metric (\ref{metR2}) to a metric with more rotation parameters. The rotation group in $(n+1)$ dimensions is $SO(n)$, and therefore the number of independent rotation
parameters is $[n/2]$. The generalized solution with $m\leq \lbrack n/2]$
rotation parameters can be written as
\begin{eqnarray}
ds^{2} &=&-\frac{r^{z}}{l^{z}}\left( \Xi dt-{{\sum_{i=1}^{m}}}a_{i}d\theta
_{i}\right) ^{2}+\frac{r^{2}}{l^{4}}{{\sum_{i=1}^{m}}}\left( a_{i}dt-\Xi
l^{2}d\theta _{i}\right) ^{2}  \notag \\
&&\ +\frac{l^2 dr^{2}}{r^{2}}-\frac{r^{2}}{l^{2}}{\sum_{i<j}^{m}}(a_{i}d\theta
_{j}-a_{j}d\theta _{i})^{2}+r^{2}\sum\limits_{i=m+1}^{n-1}d\theta _{i}^{2},
\label{metR3}
\end{eqnarray}
where $\Xi =\sqrt{1+\sum_{i}^{m}a_{i}^{2}/l^{2}}$ and the angular
coordinates are in the range $0\leq \theta _{i}<2\pi $.

This generalization can be extended to the  Lovelock-Lifshitz
$k=0$ black hole since the transformation (\ref
{tth}) does not change the $r$-dependence of the metric functions and the
vector field. Hence
\begin{eqnarray}
ds^{2} &=&-\frac{r^{z}}{l^{z}}f(r)\left( \Xi dt-{{\sum_{i=1}^{k}}}%
a_{i}d\theta _{i}\right) ^{2}+\frac{r^{2}}{l^{4}}{{\sum_{i=1}^{k}}}\left(
a_{i}dt-\Xi l^{2}d\theta _{i}\right) ^{2}  \notag \\
&&\ +\frac{l^2 dr^{2}}{r^{2}g(r)}-\frac{r^{2}}{l^{2}}{\sum_{i<j}^{k}}%
(a_{i}d\theta _{j}-a_{j}d\theta
_{i})^{2}+r^{2}\sum\limits_{i=k+1}^{n-1}d\theta _{i}^{2},  \label{metR4}
\end{eqnarray}
with the vector field
\begin{equation}
A=q\frac{r^{z}}{l^{z}}h(r)\left( \Xi dt-{{\sum_{i=1}^{k}}}a_{i}d\theta
_{i}\right) ,
\end{equation}
is a rotating black hole solution to
the field equations provided $f(r)$, $g(r)$ and $h(r)$ are chosen
to be the functions calculated in section \ref{numsol} for $k=0$.

\section{Concluding Remarks}

It is known that Einstein gravity in the presence of a massive vector field can support Lifshitz solutions \cite{Kach}.   We have demonstrated here that such solutions exist in pure Lovelock gravity provided the coupling parameters are chosen appropriately. The higher curvature terms appear to play the role of some kind of matter field,  whose nature depends on the constants of the theory.
We also found for any value of $z$ an exact vacuum asymptotically Lifshitz solution (\ref{exa}) for $k=\pm 1$, which is a black hole for $k=-1$, and a naked singularity for $k=1$.  This solution exists in both Gauss-Bonnet and 3rd order Lovelock gravity, depending on the choice of coupling.

We also obtained a broad class of solutions for Lovelock gravity coupled to a massive vector field.
After demonstrating that Lovelock gravity can support a
Lifshitz solution in the presence of a massive vector field, we searched for asymptotic
Lifshitz black holes in the presence of a massive vector field. Our numerically obtained solutions
generalize  those   obtained in Einsteinian gravity coupled to
a massive vector field  \cite{Mann,Peet,Dan}. We found that asymptotic Lifshitz solutions ($z>1$) are more sensitive to the corrections induced by Lovelock gravity than are their $z=1$ counterparts.

We also considered the thermodynamics of the black hole solutions. We found that, as in
the case of asymptotically AdS black holes of Lovelock gravity, one can have
an extreme black hole only for $k=-1$. However the  horizon radius
of both the extreme black holes of Einstein and Lovelock gravity for $k=-1$ with $z=2$ are smaller than their  $z=1$ counterparts. That is, the radius of the extreme black holes decreases as $z$ increases both in Einstein and Lovelock gravity. Also, Lovelock terms decrease the radius of
extremal black holes compared to their Einsteinian counterparts.
The temperature of a Lovelock-Lifshitz black hole with entropy $S$ is smaller than
the temperature of a Lifshitz black hole in Einstein gravity with the same entropy.
We also found, numerically,
that as $z$ increases the temperature of a black hole with a fixed
horizon radius increases. Indeed, the temperature is proportional to
$r_0^z$ for black holes with zero curvature horizon and also for large
black holes with nonzero horizon curvature. In these two cases the temperature
is proportional to $S^{z/(n-1)}$.

We investigated only the cases with $z\geq 2$
and found that these solutions are thermodynamically stable. This fact can be seen by
diagrams of temperature versus the entropy. This feature of asymptotically
Lifshitz black holes is different from the asymptotically AdS solutions that can have an
unstable phase.

Our results demonstrate that higher-order curvature corrections to ``Lifshitz holography" are under control and have a sensible physical interpretation. However there is much remaining to explore for these kinds of black holes, including the corrections they induce in the dual boundary theory and a more detailed study of the behavior for larger $n$ and larger $z$.  The rotating $k=0$ solutions should have counterparts for
the $k=\pm 1$ cases.   Work on these areas is in progress.

\section{Appendix}

Here, we investigate the behavior of the metric functions at large $r$. For
the case of $k=0$ the powers of $1/r$ may be non-integer, and therefore we
consider it separately. For this case, we use straightforward perturbation theory, writing
\begin{eqnarray*}
f(r) &=&1+\varepsilon f_{1}(r), \\
g(r) &=&1+\varepsilon g_{1}(r), \\
h(r) &=&1+\varepsilon h_{1}(r),
\end{eqnarray*}
and then finding the field equations up to the first order in $\varepsilon $. We obtain
\begin{eqnarray*}
0 &=&2r^{2}h_{1}^{\prime \prime }+2(n+z)rh_{1}^{\prime }+zr\left(
g_{1}^{\prime }-f_{1}^{\prime }\right) +2(n-1)zg_{1}, \\
0 &=&2(z-1)rh_{1}^{\prime }+(n-1)rg_{1}^{\prime }+\left[ z(z-1)+n(n-1)\right]
g_{1}-(z-1)(n+z-1)(f_{1}-2h_{1}), \\
0 &=&2(z-1)rh_{1}^{\prime }+(n-1)rg_{1}^{\prime }+\left[
z(z-1)+n(n-1)+2(n-1)(z-1)\mathcal{B}\right] g_{1}\\
&& -(z-1)(z-n+1)(f_{1}-2h_{1})
\end{eqnarray*}
where $\mathcal{B}=(l^{4}-4\hat{\alpha}_{2}l^{2}+9\hat{\alpha}_{3})/L^{4}$.
Note that all the parameters of Lovelock gravity are in $\mathcal{B}$,
with $\mathcal{B}=1$  in Einstein gravity.
The solution of the above equations may be written as
\begin{eqnarray*}
h_{1}(r) &=&C_{1}r^{-(n+z-1)}+r^{-(n+z-1)/2}\left( C_{2}r^{-\gamma
/2}+C_{3}r^{\gamma /2}\right) , \\
f_{1}(r) &=&C_{1}F_{1}r^{-(n+z-1)}+r^{-(n+z-1)/2}\left( C_{2}F_{2}r^{-\gamma
/2}+C_{3}F_{3}r^{\gamma /2}\right) , \\
g_{1}(r) &=&C_{1}G_1 r^{-(n+z-1)}+r^{-(n+z-1)/2}\left( C_{2}G_2r^{-\gamma
/2}+C_{3}G_3r^{\gamma /2}\right) ,
\end{eqnarray*}
where  $C_{1}$, $C_{2}$ and $C_{3}$ are integration constants and
\begin{eqnarray*}
\gamma  &=&\left\{(17-8\mathcal{B})z^2-2(3n+9-8\mathcal{B})z+n^2+6n+1-8\mathcal{B}\right\}^{1/2}, \\
F_1 &=&-2\left( z-1 \right)  \left(n -z-1 \right){\mathcal{K}}^{-1},\\
F_{2} &=&\left(\mathcal{F}_1-\mathcal{F}_2\right)\left\{8z\mathcal{K}\left[(z-1)\mathcal{B}+2n+z-3\right]\right\}^{-1}, \\
F_{3} &=&\left(\mathcal{F}_1+\mathcal{F}_2\right)\left\{8z\mathcal{K}\left[(z-1)\mathcal{B}+2n+z-3\right]\right\}^{-1},\\
G_1 &=&2\left( z-1 \right)  \left(n+z-1 \right){\mathcal{K}}^{-1},\\
G_{2} &=&\left(\mathcal{G}_1+\mathcal{G}_2\right)\left\{8z\mathcal{K}\left[(z-1)\mathcal{B}+2n+z-3\right]\right\}^{-1},\\
G_{3} &=&\left(\mathcal{G}_1-\mathcal{G}_2\right)\left\{8z\mathcal{K}\left[(z-1)\mathcal{B}+2n+z-3\right]\right\}^{-1},\\
\mathcal{K} &=& (z-1)(n+z-1)\mathcal{B}+z(z-1)+n(n-1),\\
\mathcal{F}_1 &=&8(z-1)[(z-1)(n+z-1)\mathcal{B}+z(z-1)+n(n-1)]\\
&& \times[(z-1)(n+3z-3)\mathcal{B}-2z^{2}+(n+3)z+n(n-2)-1],\\
\mathcal{F}_2 &=&\gamma[n-1+(z-1)\mathcal{B}]
\Big\{8(1+\mathcal{B}%
)(z-1)^{3} \\
&& +(17n-9+8\mathcal{B})(z-1)^{2}+2(n+8)(n-1)(z-1)+n^{2}(n-1)-(n-1)\gamma ^{2}\Big\},\\
\mathcal{G}_1 &=&8(z-1)[2(z-1)\mathcal{B}-3z+3n-1][(z-1)(n+z-1)\mathcal{B}+z(z-1)+n(n-1)],\\
\mathcal{G}_2 &=&\Big\{ 8(1+\mathcal{B})(z-1)^{3}+(17n-9+8\mathcal{%
B})(z-1)^{2}\\
&& +2(n+8)(n-1)(z-1) +n^{2}(n-1)\Big\} \gamma -(n-1)\gamma ^{3}.
\end{eqnarray*}
Note that the functions $f_{1}(r)$, $g_{1}(r)$ and $%
h_{1}(r)$ reduce  to those given in ref. \cite{Peet} for $\hat{\alpha}_{2}=\hat{%
\alpha}_{3}=0$  ($\mathcal{B}=1$) and $n=3$. For arbitrary values of $\hat{\alpha}_{2}$ and $\hat{%
\alpha}_{3}$, $\gamma$ is not an integer, and therefore the only integer power of $r$
is $r^{-n-z+1}$. However in Einstein gravity $\gamma$ can be an integer. This occurs for $z=2$ with either $n=3$ or $n=6$.
For $z=2$ and $n=3$, $\gamma=4$ and $C_2=0$;  the other two powers are the same
and equal to $-4$. Of course,
the next order leading term is $r^{2(-n-z+1)}=r^{-8}$ which is a second order perturbation in $\varepsilon $. For $z=2$ and $n=6$, $\gamma=5$. The largest power of $r$ in the
large $r$ expansion is $r^{-1}$. For other $z$ or $n$, even in Einstein gravity, the only integer power
of $r$ at large $r$ for $k=0$  to   first order  in $\varepsilon $ (powers larger than $2(-n-z+1)$) is $r^{-n-z+1}$.

We next obtain the coefficients of integer powers of $r$ for $k=\pm 1$ up to the first order in $\varepsilon$.
The functions
$f(r)$, $g(r)$ and $h(r)$ at large $r$ up to the first order in $\varepsilon $ may be written as
\begin{eqnarray*}
f(r) &=&1+{{\sum_{i=1}^{2(n+z)-3}}}\frac{a_{i}}{r^i}, \\
g(r) &=&1+{{\sum_{i=1}^{2(n+z)-3}}}\frac{b_{i}}{r^i}, \\
h(r) &=&1+{{\sum_{i=1}^{2(n+z)-3}}}\frac{c_{i}}{r^i}.
\end{eqnarray*}
First, we consider the problem in Einstein gravity.
For $n=3$ and $z=2$, all the coefficients of integer powers of $r$ up to the first
order ($n\geq-7$) are proportional to $k$ except the coefficient of $r^{-4}$. These coefficients are given in\cite{Mann}.
For $z=2$ and $n=6$, the first non-vanishing term is $r^{-1}$ and all the odd and even powers of $r$ are present. It is a matter of
calculation to show that the coefficients of  first three powers ($r^{-1}$, $r^{-2}$ and $r^{-3}$) are
\begin{eqnarray*}
&& a_1=3 c_1,\hspace{2.65cm} b_1=\frac{c_1}{3},\\
&& a_2=\frac{13}{6}c_1^2+\frac{4}{5}kl^2,\hspace{1.1cm} b_2=-\frac{1}{2}c_1^2+\frac{4}{5}kl^2, \hspace{.3cm}c_2=-\frac{1}{2}c_1^2+\frac{4}{5}kl^2,\\
&& a_3=-\frac{1}{27}c_1^3+\frac{136}{45}kl^2c_1,\hspace{.1cm} b_3=\frac{7}{9}c_1^3+\frac{8}{15}kl^2c_1,
\hspace{.1cm} c_3=\frac{29}{54}c_1^3+\frac{44}{45}kl^2c_1,
\end{eqnarray*}
where $c_1$ is an arbitrary coefficient.
For other values of $n$ with $z=2$ in Einstein gravity, the nonzero coefficients are those of
$r^{-2}$, $r^{-n-z+1}$, $r^{-n-z-1}$,... $r^{-n-z+1-2m}$, where $m$ is the largest integer less than $(n+z-1)/2$
up to the first order terms in $\varepsilon $ ($i<2(n+z-1)$). Of course, the next power of $r$ is $r^{-2n-2z+2}$,
which is a second order term in our analysis and is nonzero even for $k=0$ case. We note that
for $z=n-1$ in ($n+1$)-dimensional Einstein gravity, some logarithmic terms will appear.

For arbitrary values of the Lovelock coefficients,
all the odd powers of $1/r$ vanish at large $r$ and the coefficients
of $1/r^2$ are
\begin{eqnarray*}
a_{2} &=&k\frac{zl^{2}}{B}\{3\hat{\alpha}%
_{3}[2z^{3}-2(n-5)z^{2}+3(n-4)(n-5)z-(2n^{2}-11n+16)] \\
&&-2l^{2}\hat{\alpha}%
_{2}[z^{3}+(n-5)z^{2}+(n^{2}-7n+13)^{2}z-(2n^{2}-10n+13)]+(n-2)[(n-3)z-2n+5]l^{4}\},
\\
b_{2} &=&k\frac{l^{2}}{B}\{3\hat{\alpha}%
_{3}[2z^{3}+2(n-5)z^{2}+(n^{2}-10n+22)z-2(n-3)^{2}]+(n-2)[(n-4)z-2(n-3)]l^{4} \\
&&-2l^{2}\hat{\alpha}%
_{2}[z^{3}+(n-5)z^{2}+(n-3)(n-5))z-(2n^{2}-11n+15)]\},
\\
c_{2} &=&k\frac{zl^{2}}{2B}\{3\hat{\alpha}%
_{3}[2z^{3}+2(n-4)z^{2}+(n^{2}-9n+16)z-2(n-2)(n-3)] \\
&&-2l^{2}\hat{\alpha}%
_{2}[z^{3}-(n-4)z^{2}+(n^{2}-7n+11)z-(2n^{2}-9n+10)]-(n-2)[(n-3)z-2(n-2)]l^{4}\},
\end{eqnarray*}
where
\begin{equation*}
B=(z+n-3)\left\{ 3\hat{\alpha}_{3}(z-2)(z+n-3)-2\hat{\alpha}%
_{2}l^{2}[z(n-3)-2n+5]-\left[ z^{2}-(n-1)z+2(n-2)\right] l^{4}\right\} .
\end{equation*}
The higher order coefficients with even power of $r$ ($r^{-4}$, $r^{-6}$,..) are present and can be easily calculated but we shall not write
them here as they are quite lengthy.  In this case all $a_{n}$'s are proportional to $k$.
Thus, this feature is again a difference between Einstein and Lovelock gravity. That is,
in Lovelock gravity all the even powers of $r$ up to $r^{-n-z+1}$ are
present in the expansion of the functions at large $r$, while in Einstein gravity this does not occur
for $z=2$.

\section*{Acknowledgements}
This work was supported by the Natural Sciences and Engineering Research Council of Canada and
Research Council of Shiraz University

\end{document}